\documentclass[a4paper,fleqn,usenatbib]{mnras}

\usepackage{newtxtext,newtxmath}
\usepackage[T1]{fontenc}
\usepackage{ae,aecompl}

\usepackage{graphicx}	% Including figure files
\usepackage{amsmath}	% Advanced maths commands
\usepackage{amssymb}	% Extra maths symbols

\title[DRAGONS: Clustering of high redshift galaxies]{Dark-ages reionization and galaxy formation simulation XI: Clustering and halo masses of high redshift galaxies}

\author[Park et al.]{Jaehong~Park$^{1,3}$\thanks{jaehong.park@sns.it}, Han-Seek~Kim$^{1}$\thanks{hansikk@unimelb.edu.au}, Chuanwu Liu$^{1}$,  M. Trenti$^{1}$, Alan R. Duffy$^{2}$, \\ \vspace{0.3mm}\\
      {\LARGE {\rm Paul M. Geil$^{1}$, Simon J. Mutch$^{1}$, Gregory B. Poole$^{1}$, Andrei Mesinger$^{3}$ and}}\\ \vspace{0.3mm}\\
       {\LARGE {\rm J. Stuart B.~Wyithe$^{1}$}}\\
       $^1$School of Physics, The University of Melbourne, Parkville, VIC 3010, Australia\\
       $^2$Centre for Astrophysics and Supercomputing, Swinburne University of Technology, PO Box 218, Hawthorn, VIC 3122, Australia\\
       $^3$Scuola Normale Superiore, Piazza dei Cavalieri 7, I-56126 Pisa, Italy\\
	}
\date{Accepted XXX. Received YYY; in original form ZZZ}

\pubyear{2017}

\begin{document}
\label{firstpage}
\pagerange{\pageref{firstpage}--\pageref{lastpage}}
\maketitle

%%%%%%%%%%%%%%%%%%%%%%%%%%%%%%%%%%%%%%%%%%%%%%%%%%%%%%%
%                                                                                                                                                                                                   %
%  Abstract                                                                                                                                                                                  %
%                                                                                                                                                                                                   
%%%%%%%%%%%%%%%%%%%%%%%%%%%%%%%%%%%%%%%%%%%%%%%%%%%%%%%
\begin{abstract}
We investigate the clustering properties of Lyman-break galaxies (LBGs) at $z\sim6$ - $8$. Using the semi-analytical model {\scshape Meraxes} constructed as part of the Dark-ages Reionization And Galaxy-formation Observables from Numerical Simulation (DRAGONS) project, we predict the angular correlation function (ACF) of LBGs at $z\sim6$\,-\,$8$. Overall, we find that the predicted ACFs are in good agreement with recent measurements at $z\sim 6$ and $z\sim 7.2$ from observations consisting of the Hubble eXtreme Deep Field (XDF), the Hubble Ultra-Deep Field (HUDF) and Cosmic Assembly Near-infrared Deep Extragalactic Legacy Survey (CANDELS) field. We confirm the dependence of clustering on luminosity, with more massive dark matter haloes hosting brighter galaxies, remains valid at high redshift. The predicted galaxy bias at fixed luminosity is found to increase with redshift, in agreement with observations. We find that LBGs of magnitude $M_{{\rm AB(1600)}} < -19.4$ at $6\lesssim z \lesssim 8$ reside in dark matter haloes of mean mass $\sim 10^{11.0}$\,-\,$10^{11.5}\,M_{\rm \odot}$, and this dark matter halo mass does not evolve significantly during reionisation.
\end{abstract}


%%%%%%%%%%%%%%%%%%%%%%%%%%%%%%%%%%%%%%%%%%%%%%%%%%

%%%%%%%%%%%%%%%%% BODY OF PAPER %%%%%%%%%%%%%%%%%%
%%%%%%%%%%%%%%%%%%%%%%%%%%%%%%%%%%%%%%%%%%%%%%%%%%%%%%%
%                                                                                                                                                                                                   %
%  Introduction                                                                                                                         %
%                                                                                                                                                                                                   %
%%%%%%%%%%%%%%%%%%%%%%%%%%%%%%%%%%%%%%%%%%%%%%%%%%%%%%%
\section{Introduction}\label{sec:intro}
Clustering of galaxies allows us to probe the large-scale structure of the Universe as a biased tracer of the density field, and galaxy formation physics by providing a measure of dark matter halo mass. The formation and evolution of dark matter haloes are described by analytic models \citep[e.g.][]{MW96,Cooray2002} and by N-body simulations \citep[e.g.][]{Springel2005}. Galaxies are thought to form inside those dark matter haloes \citep{White&Frenk1991,Cole1991}, but we are still from having a complete understanding of the galaxy formation process, which involves complicated non-linear physical processes \citep[e.g.][]{Baugh2006, Benson2010,Schaye2015}.

Accurate measurements of the clustering of galaxies in the local Universe have determined the clustering dependence on luminosity \citep[e.g.][]{Norberg2001,Norberg2002a,Zehavi2005,Zehavi2011}. These observational results provide strong constraints on theoretical predictions for galaxy properties \citep[e.g.][]{Henriques2015,Lacey2015,Schaye2015}.

At higher redshift, galaxies selected by the Lyman-break technique (LBGs) are the most extensively studied sources \citep[e.g.][]{Giavalisco2002}. Galaxies at a specific redshift can be selected using a colour selection criteria designed to detect spectral features of star-forming galaxies corresponding to absorption of the rest-frame far-UV emission (below $1216\, {\rm \AA}$) by neutral hydrogen. Since the work of \cite{Steidel&Hamilton1993} and \cite{Steidel1996} at $z\sim3$, this technique has proved an effective and efficient method of discovery. This technique has now been extended to detect galaxies up to $z\sim10$ \citep[e.g.][]{Oesch2012,Bradley2012,McLure2013,Zitrin2014,Duncan2014,Bouwens2015,Bouwens2016}, providing the dominant source of information about galaxies during reionisation.

Over the past decade, clustering of LBGs has been measured in the redshift range $z\sim3$\,-\,7~\citep{Ouchi2005,Cooray2006,Kashikawa2006,Lee2006,McLure2009,Hildebrandt2009,Harikane2015}. Recently, \cite{Bouwens2015} identified LBGs in the redshift range $z\sim4-10$ in a combined survey field consisting of the Hubble eXtreme Deep Field (XDF), the Hubble Ultra-Deep Field (HUDF) and Cosmic Assembly Near-infrared Deep Extragalactic Legacy Survey (CANDELS) field. Using the sample of \cite{Bouwens2015}, \cite{Rob2014} measured the angular correlation function (ACF) of LBGs in the redshift range $z\sim4$\,-\,7.2. This includes the first measurement of LBG clustering at $z\sim7.2$, using combined samples with $z\sim7$ LBGs and $z\sim8$ LBGs. Since \cite{Rob2014} measured the ACF in each survey field independently, this measurement gives us an estimate of sample variance by comparing the results from different fields, while the measured ACF from the Hubble eXtreme Deep Field (XDF) allows us to investigate the clustering of the fainter LBGs. More recently, \cite{Harikane2015} measured the ACF of LBGs in the redshift range $z\sim4$\,-\,7 using a combined data set from HUDF and CANDELS fields, and from XMM and GAMA09h fields of Subaru/Hyper Suprime-Cam (HSC) observations. They estimated the dark matter halo mass using Halo Occupation Distribution (HOD) modelling.

Recently, based on the development of observations, semi-analytical models \citep[e.g.][]{Samui2007, Samui2014,Jose2014,Lacey2015,Liu2016, Jose2017} and hydrodynamic simulations \citep[e.g.][]{Ocvirk2016,Waters2016} predict UV luminosity functions of LBGs up to $z\sim6$\,-\,$10$. Although the predicted luminosity functions provide the abundance of galaxies, they do not show the spatial distribution of galaxies. Comparing the model prediction of clustering with observations therefore provides additional insight into high redshift galaxies. Previous studies have been restricted to comparison of the predictions from theoretical models with observational measurements \citep[e.g.][]{Kashikawa2006,Jose2013} at redshifts less than $z\sim5$ due to the insufficient number of LBGs observed at $z>5$. Recently, \cite{Waters2016} predicted the galaxy bias from the real-space correlation function at higher redshifts of $z=8$\,-\,10 but did not compare their result with measured ACFs.

In this paper, we investigate the clustering properties of LBGs at $z\sim6$\,-\,8. We use the semi-analytical galaxy formation model, {\scshape Meraxes} \citep{Mutch2016}, which is designed to study galaxy formation during the Epoch of Reionisation (EoR). {\scshape Meraxes} is able to describe the luminosity function, stellar mass function, and their evolution \citep{Liu2016}. We predict the ACF of LBGs selected from the model, providing model predictions for clustering of LBGs up to $z\sim8$. We compare these model predictions with the clustering measured from \cite{Rob2014}, and also compare with the results from \cite{Harikane2015}.

We begin in Section~\ref{sec:model} by briefly describing {\scshape Meraxes}. In Section~\ref{LBGs}, we present the methodology used to select LBGs, show a resulting luminosity function, and describe how to compute ACFs in the model. We present the predictions for the clustering properties and compare the model predictions with observations in Section~\ref{sec:comparison}. Then, we conclude in Section~\ref{sec:conclusion}. Throughout the paper we use apparent magnitudes in the observers frame in the AB system. Where we refer to the UV magnitude, this corresponds to the rest-frame $1600{\rm \AA}$ AB magnitude. We employ a standard spatially-flat ${\rm \Lambda}$CDM cosmology based on {\it Planck} 2015 result \citep{Planck2015}: ($h$, $\Omega_{\rm m}$, $\Omega_{\rm b}$, $\Omega_{\Lambda}$, $\sigma_{8}$, $n_{\rm s}$)=(0.678, 0.308, 0.0484, 0.692, 0.815, 0.968).

%%%%%%%%%%%%%%%%%%%%%%%%%%%%%%%%%%%%%%%%%%%%%%%%%%%%%%%
%                                                                                                                                                                                                   %
%   Dark-ages reionization & galaxy formation simulation
%                                                                                                                                                                                                   %
%%%%%%%%%%%%%%%%%%%%%%%%%%%%%%%%%%%%%%%%%%%%%%%%%%%%%%%
\section{The model}\label{sec:model}
In this section we summarise the model used in this study. In \S\,\ref{sec:meraxes}, we briefly introduce the model {\scshape Meraxes}. Then we describe how to compute magnitudes of galaxies in \S\,\ref{sec:prediction_of_magnitudes}. For further details, interested readers are referred to \cite{Mutch2016} for {\scshape Meraxes}, and to \cite{Liu2016} for the UV luminosity function.
%%%%%%%%%%%%%%%%%%%%%%%%%%%%%%%%%%%%%%%%%%%%%%%%%%%%%%%%%%
%     The Meraxes
%%%%%%%%%%%%%%%%%%%%%%%%%%%%%%%%%%%%%%%%%%%%%%%%%%%%%%%%%
\subsection{Meraxes}\label{sec:meraxes}
{\scshape Meraxes} is the semi-analytical model \citep{Mutch2016} constructed as part of the Dark-ages Reionisation And Galaxy-formation Observables from Numerical Simulations\footnote{http://dragons.ph.unimelb.edu.au} (DRAGONS) project. DRAGONS integrates a semi-analytical model with a semi-numerical model \citep{Mesinger2011} in order to self-consistently simulate the reionisation and galaxy formation processes. 

{\scshape Meraxes} computes the formation and evolution of galaxy properties in the redshift range $z\ge 5$. {\scshape Meraxes} has been developed based on the model of \cite{Croton2006} and extended in \cite{Guo2011}. We implement {\scshape Meraxes} within the N-body dark matter simulation {\it Tiamat} (see \citealp{Poole2016} for more details of {\it Tiamat}). In the fiducial {\it Tiamat} simulation the particle mass is 3.89$\times$10$^{6}$${\rm M_{\odot}}$, with a side length of $100\,{\rm Mpc}$. {\it Tiamat} provides 100 output snapshots between $z=35-5$ with a cadence of 11.1\,Myr. This high cadence is necessary because the galaxy dynamical time at $z\gtrsim 6$ becomes comparable to the lifetime of massive stars. This allows us to compute time-resolved supernova feedback.

We describe basic differences of Meraxies to traditional semi-analytical models below (see \citealp{Mutch2016} for more details of {\scshape Meraxes}).

 (i) Merger trees: The merger trees are constructed ``horizontally''. During the EoR, ionising photons from galaxies tens of Mpc away can ionise the intergalactic medium (IGM), affecting subsequent  galaxy formation processes. {\scshape Meraxes} takes into account radiation from galaxies which are spatially associated with each other at each snapshot.%This allows {\scshape Meraxes} to simulate the reionisation process realistically.}

 (ii) Delayed supernova feedback: {\scshape Meraxes} follows the parametrisation of \cite{Guo2013} for the efficiency of the supernova feedback. However, the N-body simulation ({\it Tiamat}) on which {\scshape Meraxes} is run has one snapshot per $\sim 11.1$\,Myr, providing much higher time resolution than many semi-analytical models. This high time resolution is required at high redshift where the dynamical time in galaxies becomes smaller than the lifetime of massive stars. The life time of the massive stars corresponds to $\sim4$ snapshots in this model, and {\scshape Meraxes} adopts a delayed supernova feedback scheme. 
The total amount of supernova energy released by a galaxy at a snapshot is computed by tracking the total mass of stars formed in each galaxy for last $40\,{\rm Myr}$.

 (iii) Reionisation: To model the reionisation process, {\scshape Meraxes} uses a modified 21{\scshape cmfast} \citep{Mesinger2011,Sobacchi2013} algorithm. The criterion to find ionised regions can be written as 
%===== Equation =====
\begin{equation}\label{eq:reion2}
\xi\frac{m_{\ast}(r)}{M_{\rm tot}} \ge 1,
\end{equation}
%==================
where $m_{\ast}(r)$ is the integrated stellar mass within radius $r$, $M_{\rm tot}$ is the total mass within $r$, and $\xi$ is an \ion{H}{ii} ionising efficiency.

From the equation~(\ref{eq:reion2}), {\scshape Meraxes} computes a global ionisation structure for the simulation volume. Then, {\scshape Meraxes} computes the value of the baryon fraction modifier, $f_{\rm mod}$,  \citep{Sobacchi2013}
%===== Equation =====
\begin{equation}\label{eq:reion4}
f_{\rm mod} = 2^{-M_{\rm filt}/M_{\rm vir}},
\end{equation}
%==================
where $M_{\rm filt}$ is the filtering mass at $f_{\rm mod}=0.5$, given by
%===== Equation =====
\begin{equation}\label{eq:reion5}
M_{\rm filt} = M_0J_{21}^a \left( \frac{1+z}{10}\right)^b \left[ 1 - \left(\frac{1+z}{1+z_{\rm ion}}\right)^c\right]^d,
\end{equation}
%==================
where $J_{21}$ is the local ionising intensity and $z_{\rm ion}$ is the redshift at which a halo was first ionised. The values of parameters are ($M_0$, $a$, $b$, $c$, $d$)$=$($2.8\times 10^9\,{\rm M_{\odot}}$, $0.17$, $-2.1$, $2.0$, $2.5$) as found by \cite{Sobacchi2013}. Based on the ionised structure and $z_{\rm ion}$, each halo uses the baryon fraction modifier, $f_{\rm mod}$, to compute the infalling baryonic mass.

 (iv) Baryonic infall: Ionising UV background radiation heats the IGM and raises the local Jeans mass, reducing the fraction of the baryonic infall, $f_{\rm b}$ \citep{dijkstra2004}. {\scshape Meraxes} parameterises this reduction using a baryon fraction modifier, $f_{\rm mod}$. The infalling baryonic mass into the dark matter haloes is 
%===== Equation =====
\begin{equation}\label{eq:infall_mass}
m_{\rm infall} = f_{\rm mod}f_{\rm b}M_{\rm vir} - \sum_{i=0}^{N_{\rm gal}-1} m_{\ast}^i + m_{\rm cold}^i + m_{\rm hot}^i + m_{\rm ejected}^i,
\end{equation}
%==================  
where $M_{\rm vir}$ is the mass of the halo, $N_{\rm gal}$ is the number of galaxies in the dark matter halo, and the baryon fraction modifier has a range, $0 \le f_{\rm mod} \le 1$. $m_{\ast}$, $m_{\rm cold}$, $m_{\rm hot}$, and $m_{\rm ejected}$ are the stellar mass, cold gas mass, hot gas mass and the ejected gas mass from the dark matter halo, respectively.

%---------------------------------------------------------------------------------------------------------------------------------------------------%
% Figure : Redshift distribution N(z)
%---------------------------------------------------------------------------------------------------------------------------------------------------%
\begin{figure*}
\begin{center}
\includegraphics[width=5.5cm]{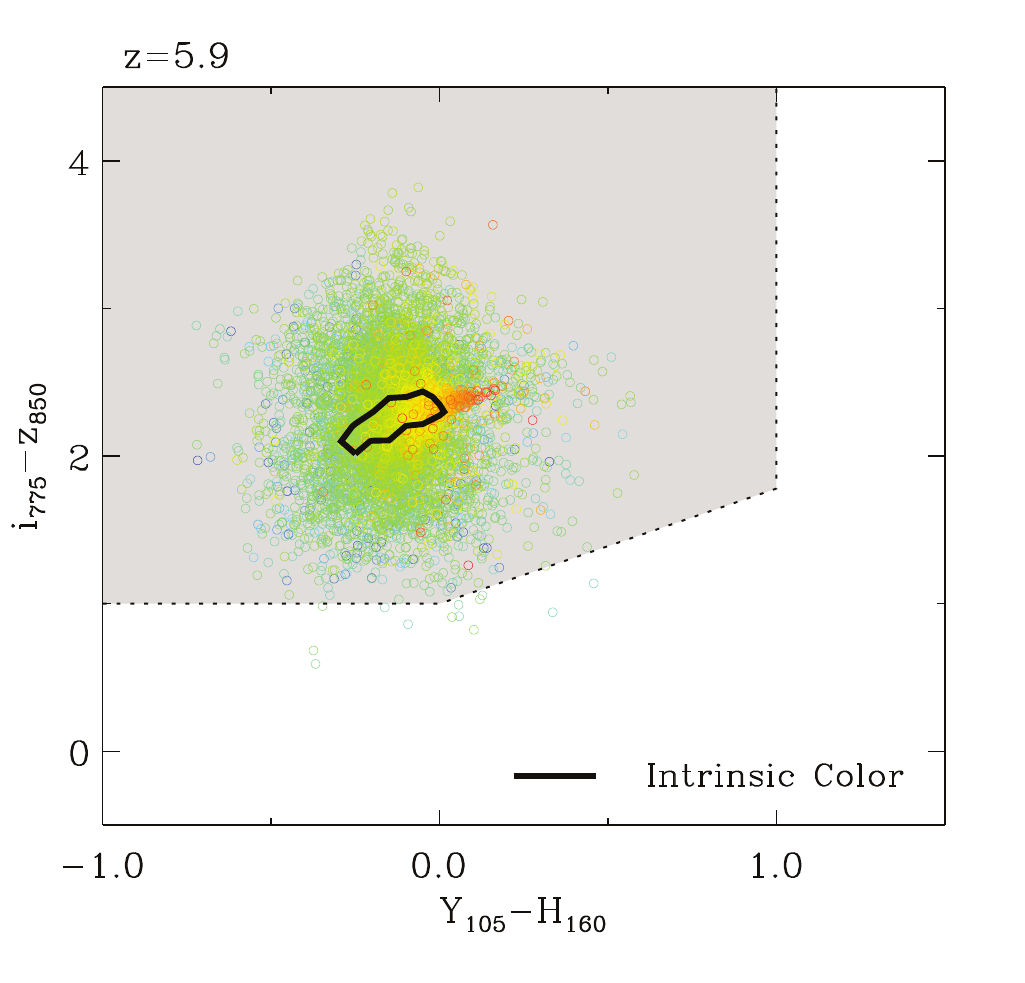}
\includegraphics[width=5.5cm]{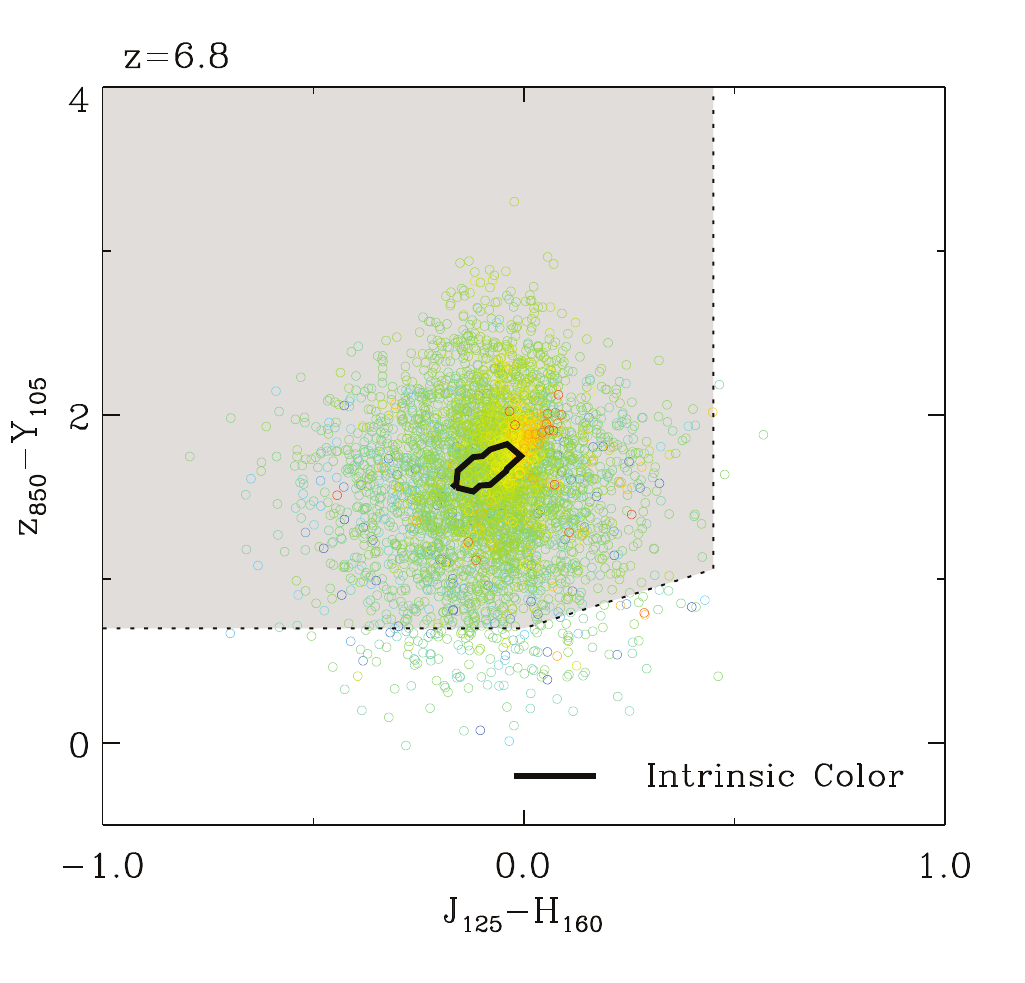}
\includegraphics[width=6.3cm]{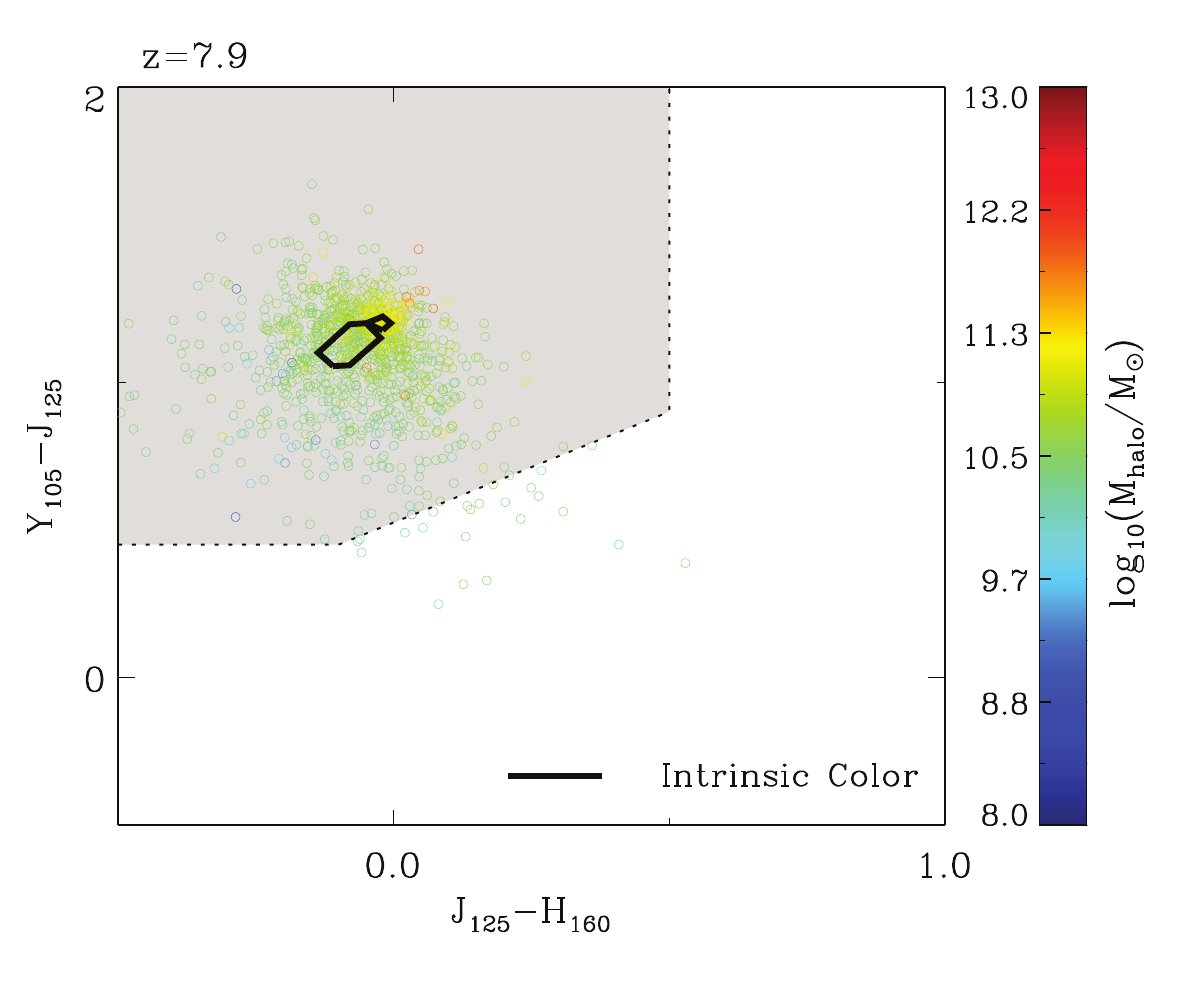}
\end{center}
\vspace{-3mm}
\caption{
Predicted colour-colour diagrams showing colours of model galaxies at mean redshifts of 5.9 (left), 6.8 (central) and 7.9 (right), corresponding to the colour selection criteria. Different colour symbols correspond to the host dark matter halo mass as indicated by the colour bar. In each panel, thick contours show the distribution of intrinsic colours, i.e. without photometric errors, at the mean redshifts. All contours enclose 99.4 per cent of model galaxies. Grey shaded regions and dotted lines represent the colour selection criteria adopted from \protect\cite{Bouwens2015}. The panels show colours of model galaxies brighter than the XDF flux limit. Compared to the predicted distribution of intrinsic colours, we can see the noticeable scatter for the colours of model galaxies.
}
\label{fig:CCD}
\end{figure*}
%---------------------------------------------------------------------------------------------------------------------------------------------------%
%%%%%%%%%%%%%%%%%%%%%%%%%%%%%%%%%%%%%%%%%%%%%%%%%%%%%%%%%%
%     Prediction of magnitudes
%%%%%%%%%%%%%%%%%%%%%%%%%%%%%%%%%%%%%%%%%%%%%%%%%%%%%%%%%
\subsection{Prediction properties of model galaxies}\label{sec:prediction_of_magnitudes}
\subsubsection{Lyman $\alpha$ absorption}\label{sec:ly-alpha_absorption}
To compute the intrinsic luminosity of each model galaxy, we start by building a star-formation history as a function of time by tracing all progenitors of a star, and calculating the intrinsic stellar luminosity. We use STARBURST99 \citep{Leitherer1999,Leitherer2014} to model stellar energy distributions (SED), following the same methodology of \cite{Liu2016}. The UV radiation from galaxies is absorbed by neutral hydrogen in the intergalactic medium. We calculate this attenuation by adopting an effective ${\rm Ly}\,\alpha$ absorption optical depth from \cite{Fan2006}. \cite{Fan2006} found that the effective optical depth evolves rapidly, $\tau_{\rm eff} \propto (1+z)^{10.9}$, at $z= 5.5$\,-\,$6.3$. For simplicity, we use this relation for all redshifts at $z\geq 5.5$. Since the observed ${\rm Ly}\,\alpha$ flux vanishes at $z>6$, this assumption does not affect LBG selections (see \cite{Liu2016} for more details).

%The magnitudes of galaxies are also attenuated by dust in the interstellar medium. The amount of dust attenuation at $1600\,{\rm \AA}$ can be expressed using the observed UV continuum slope \citep{Meurer1999}. Adopting the assumption that galaxies have similar intrinsic UV continuum slopes \citep[e.g.][]{Leitherer1995} and values of the observed UV continuum slope taken from observations \citep{Bouwens2014}, we calculate the amount of dust attenuation at $1600\,{\rm \AA}$. Then, by applying a reddening curve derived by \cite{Calzetti2000} to the predicted SED, we obtain magnitudes of model galaxies including the effect from dust attenuation.

\subsubsection{Dust attenuation}\label{sec:dust_attenuation}
Following the discussion in \cite{Liu2016} the rest-frame UV continuum for a galaxy can be written as
%======= Equation: Magnitude with error
\begin{equation}\label{eq:UV-continuum}
f_{\lambda} \propto \lambda^{\beta},
\end{equation}
%==============================
where $f_{\lambda}$ is the flux density per wavelength interval and $\beta$ is the UV continuum slope. Since the amount of dust attenuation increases with shorter wavelengths, the dust-attenuation makes the continuum slope steepen. UV flux attenuated by dust grains within galaxies can be parametrised as%By the continuum slope, the dust attenuation can be quantified. 
%======= Equation: UV flux density
\begin{equation}\label{eq:UV-flux}
F_{\rm o}(\lambda) = F_{\rm i}10^{-0.4\,A_{\lambda}},
\end{equation}
%==============================
where $F_{\rm o}$ and $F_{\rm i}$ are the observed and intrinsic continuum flux densities, and $A_{\lambda}$ is the change in magnitude at rest-frame wavelength $\lambda$. To compute the dust-attenuated UV continuum slope, $\beta$, we use the relation between the observed UV continuum, $\beta$, and the UV dust attenuation \citep{Meurer1999}
%======= Equation: A_1600 vs beta
\begin{equation}\label{eq:relation-Meurer1999}
A_{1600}=4.43 + 1.99\,{\beta},
\end{equation}
%==============================
where $A_{1600}$ is the dust attenuation at $1600\,{\rm \AA}$. The values of $\beta$ can be obtained from observations. \cite{Bouwens2014} found a piece-wise linear relation between the mean of $\beta$ and $M_{\rm AB,1600}$ at $z\sim4$\,-\,$6$
%======= Equation: beta vs Muv
\begin{equation} \label{eq:beta_vs_Muv1}
	\beta = \left\{ \begin{array}{ll}
          \frac{{\rm d}\beta}{{\rm d}M_{\rm AB,1600}}(M_{\rm AB,1600}+18.8 ) + \beta_{M_{\rm AB,1600}=-18.8},\\
           \qquad\qquad\qquad\qquad\qquad M_{\rm AB,1600}\leq 18.8, \\
          -0.08(M_{\rm AB,1600}+18.8 ) + \beta_{M_{\rm AB,1600}=-18.8}, \\
           \qquad\qquad\qquad\qquad\qquad M_{\rm AB,1600} > 18.8,
       \end{array} \right.
\end{equation} 
%==============================
where ${\rm d}\beta/{\rm d}M_{\rm AB,1600}$ and $\beta_{M_{\rm AB,1600}=-18.8}$ are taken from table 4 of \cite{Bouwens2014}. For galaxies at $z\sim7$\,-\,$8$, they also found a linear relation
%======= Equation: beta vs Muv
\begin{equation}\label{eq:beta_vs_Muv2}
\beta = \frac{{\rm d}\beta}{{\rm d}M_{\rm AB,1600}} (M_{\rm AB,1600} + 19.5) + \beta_{M_{\rm AB,1600}=-19.5},
\end{equation}
%==============================
where ${\rm d}\beta/{\rm d}M_{\rm AB,1600}$ and $\beta_{M_{\rm AB,1600}=-19.5}$ are taken from table 3 of \cite{Bouwens2014}. We assume the values of $\beta$ are distributed with Gaussian standard deviation of 0.35 \citep{Bouwens2014}. Using equation\,(\ref{eq:relation-Meurer1999}) we obtain the relation between the mean dust attenuation, $\left< A_{1600} \right>$, and the intrinsic UV luminosity at $1600\,{\rm \AA}$. These luminosities are converted to the intrinsic rest-frame magnitude,  $M^{i}_{\rm AB,1600}$, using SED with tophat bands of $100\,{\rm \AA}$ bandwidth.

At other wavelengths the dust attenuation can be written as
%======= Equation: A_lambda
\begin{equation}\label{eq:A_lambda}
A_{\lambda}=E(B-V)k(\lambda),
\end{equation}
%==============================
where $E(B-V)$ is the colour excess. The dust reddening curve $k(\lambda)$ is given by \cite{Calzetti2000}
%======= Equation: Reddening curve
\begin{equation}\label{eq:reddening_curve}
	\beta = \left\{ \begin{array}{ll}
          2.659\left( -2.156 + \frac{1.509}{\lambda} - \frac{0.198}{\lambda^2} + \frac{0.011}{\lambda^3}\right) + R_{V},\\
           \qquad\qquad\qquad\qquad\qquad 0.12\,{\rm \mu m} \leq \lambda < 0.63\,{\rm \mu m}, \\
          2.659\left( -1.857 + \frac{1.040}{\lambda} \right) + R_{V}, \\
           \qquad\qquad\qquad\qquad\qquad 0.63\,{\rm \mu m} \leq \lambda < 2.20\,{\rm \mu m},
       \end{array} \right.
\end{equation} 
%==============================
where $\lambda$ is the rest-frame wavelength in units of ${\rm \mu m}$ and $R_{V} = 4.05 \pm 0.80$ is the effective obscuration. For wavelength $\lambda < 0.12\,{\rm \mu m}$, we extrapolate the reddening curve.

%---------------------------------------------------------------------------------------------------------------------------------------------------%
% Figure : Redshift distribution N(z)
%---------------------------------------------------------------------------------------------------------------------------------------------------%
\begin{figure*}
\begin{center}
\includegraphics[width=18cm]{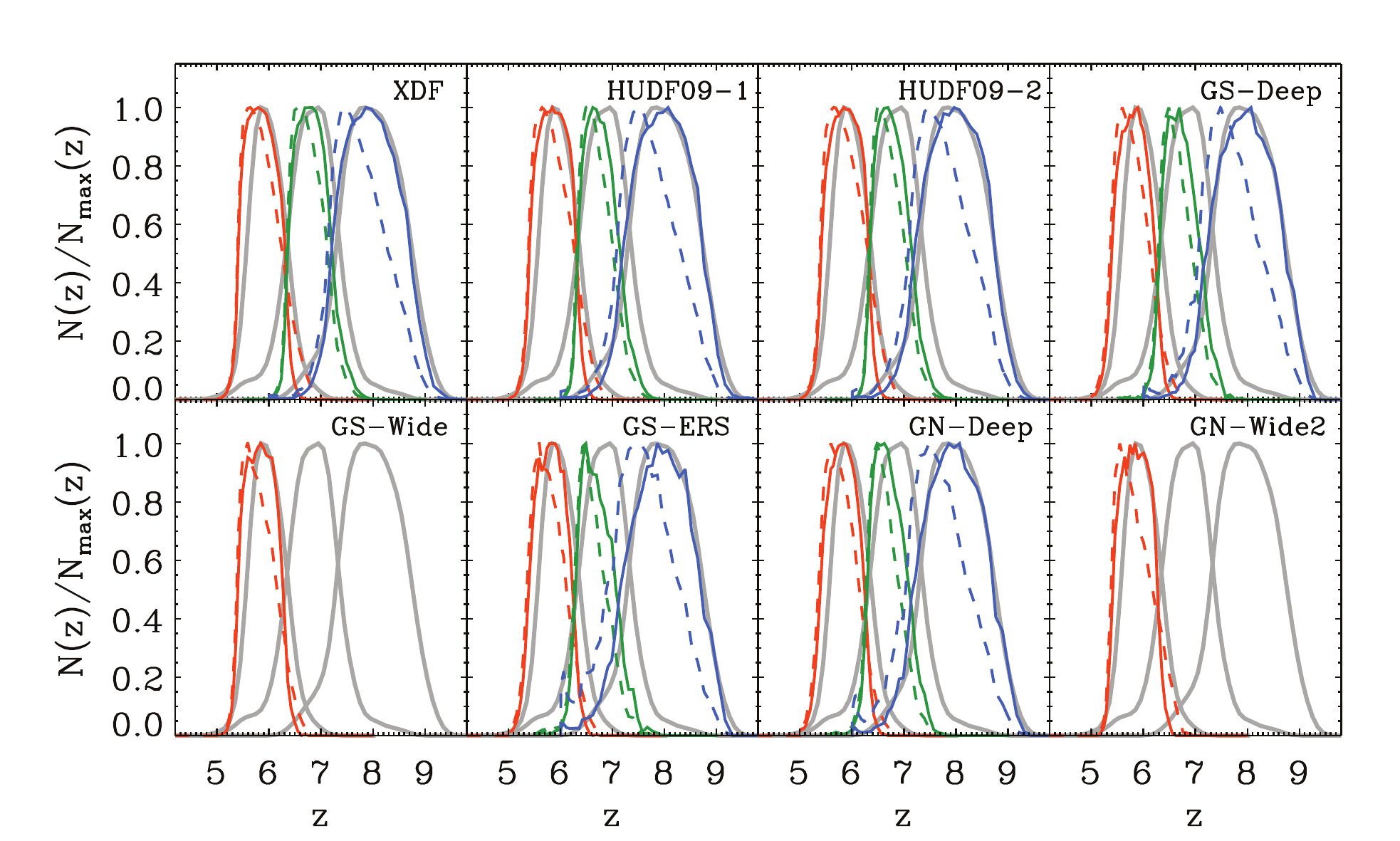}
\end{center}
\vspace{-3mm}
\caption{
Redshift distributions of selected LBGs for different observational flux limits of the XDF and CANDELS survey fields, as labelled in each panel. The redshift distributions are normalised to have a maximum value of unity. Solid lines represent the redshift distribution using the ratio of model galaxies selected as LBGs to all model galaxies at a given magnitude bin in each snapshot, and dashed lines represent the redshift distribution using equation\,(\ref{eq:Nz}). Grey lines represent the redshift distribution of oserved LBGs estimated by \protect\cite{Bouwens2015} using Monte-Carlo simulations. Note that we do not plot the predicted distribution at $z\sim7$ and 8, since LBGs are not identified in GS-Wide and GN-Wide2 fields at $z\sim7.2$ \protect\citep{Rob2014}.
}
\label{fig:Nz6}
\end{figure*}
%---------------------------------------------------------------------------------------------------------------------------------------------------%

%%%%%%%%%%%%%%%%%%%%%%%%%%%%%%%%%%%%%%%%%%%%%%%%%%%%%%%%%%
%
%     Lyman-break galaxies in the model
%
%%%%%%%%%%%%%%%%%%%%%%%%%%%%%%%%%%%%%%%%%%%%%%%%%%%%%%%%%
\section{Lyman-break galaxies in the model}\label{LBGs}

In this section we describe how we select model Lyman-break galaxies (LBGs). Then, we compare the predicted results with observations.
%%%%%%%%%%%%%%%%%%%%%%%%%%%%%%%%%%%%%%%%%%%%%%%%%%%%%%%%%%
%     Selecting Lyman-break galaxies
%%%%%%%%%%%%%%%%%%%%%%%%%%%%%%%%%%%%%%%%%%%%%%%%%%%%%%%%%
\subsection{Selecting Lyman-break galaxies}\label{selecting_LBGs}
To select model LBGs we use a similar method to that described in \cite{Liu2016}. They applied the colour selection criteria from \cite{Bouwens2015} to galaxies generated from Meraxes. The selection criteria for LBGs at $z\sim6$ are
\begin{gather}\label{colour-selection:z6}
(i_{775}-z_{850} >1.0) \ \wedge \ (Y_{105}-H_{160} <1.0) \ \wedge \nonumber \\ 
       (i_{775}-z_{850}) >0.78(Y_{105}-H_{160})+1.0, \\
       (\text{not in } z\sim7 \text{ selection}),\nonumber
\end{gather}
for LBGs at $z\sim7$ are 
\begin{gather}\label{colour-selection:z7}
(z_{850}-Y_{105} >0.7) \ \wedge \ (J_{125}-H_{160} <0.45) \ \wedge  \nonumber\\
       (z_{850}-Y_{105} > 0.8(J_{125}-H_{160})+0.7),\\
       (\text{not in } z\sim8 \text{ selection}),\nonumber \nonumber
\end{gather}
and for LBGs at $z\sim8$ are 
\begin{gather}\label{colour-selection:z8}
(Y_{125}-J_{125} >0.45) \ \wedge \ (J_{125}-H_{160} <0.5) \ \wedge  \\
       (Y_{105}-J_{125} > 0.75(J_{125}-H_{160})+0.525),\nonumber
\end{gather}
where $\wedge$ represents the logical AND symbol. $i_{775}$, $z_{850}$, $Y_{105}$, $J_{125}$ and $H_{160}$ correspond to the magnitudes of F775W, F850LP, F105W, F125W and F160W bands in ACS and WFC3/IR, respectively. Together with the above criteria, \cite{Bouwens2015} adopted additional criteria to exclude low redshift interlopers. We checked that the additional criteria do not change results of selecting model LBGs. 

%----------------------------------------------------------------
% Table: flux limits
%----------------------------------------------------------------
\begin{table*}%[!hbp]
\begin{center}
\caption{
         Flux limits and areas of the individual survey fields. Each magnitude limit is quoted as a 5\,$\sigma$ depth \protect\citep{Bouwens2015}. The units of area is ${\rm arcsec^2}$. The last two columns represent the number of LBGs at each redshift \protect\citep{Rob2014}.
                  }
\begin{tabular} {ccccccccc}
 \\
 \hline\\[-3.0mm]
 Field              &         Area  &  $i_{775}$  &  $z_{850}$  &  $Y_{105}$  &  $J_{125}$  &  $H_{160}$  & $z\sim 6$  &  $z\sim 7.2$ \\[0.5mm] \hline\\[-2.5mm]
 XDF               &         4.7    &      29.8      &     29.2        &       29.7       &       29.3       &      29.4        &    104        &      149        \\[0.5mm]
 HUDF09-1     &         4.7    &      28.5      &      28.4       &       28.3        &      28.5       &      28.3        &     38         &        52        \\[0.5mm]
 HUDF09-2     &         4.7    &      28.8      &     28.8        &       28.6        &      28.9       &      28.7        &      36         &       54         \\[0.5mm]
 GS-Deep       &         64.5  &      27.5       &     27.3        &       27.5        &      27.8       &      27.5        &    203        &      134         \\[0.5mm]
 GS-Wide        &        34.2  &     27.5       &      27.1        &      27.0        &      27.1       &      26.8         &       41       &        .          \\[0.5mm]  
 GS-ERS        &         40.2  &     27.2      &      27.1        &      27.0        &      27.6       &      27.4          &     62         &         64         \\[0.5mm] 
 GN-Deep       &         62.9  &      27.3      &      27.3        &      27.3        &      27.7       &      27.5         &    197        &       220        \\[0.5mm] 
 GN-Wide2      &        60.9  &      27.2     &      27.2        &      26.2        &       26.8      &      26.7          &      51        &        .       \\[0.5mm] 
  \hline
\end{tabular}
\label{Table:Flux_limit}
\end{center}
\end{table*}

%---------------------------------------------------------------------------------------------------------------------------------------------------%
% Figure : Luminosity function
%---------------------------------------------------------------------------------------------------------------------------------------------------%
\begin{figure*}
\begin{center}
\includegraphics[width=18cm]{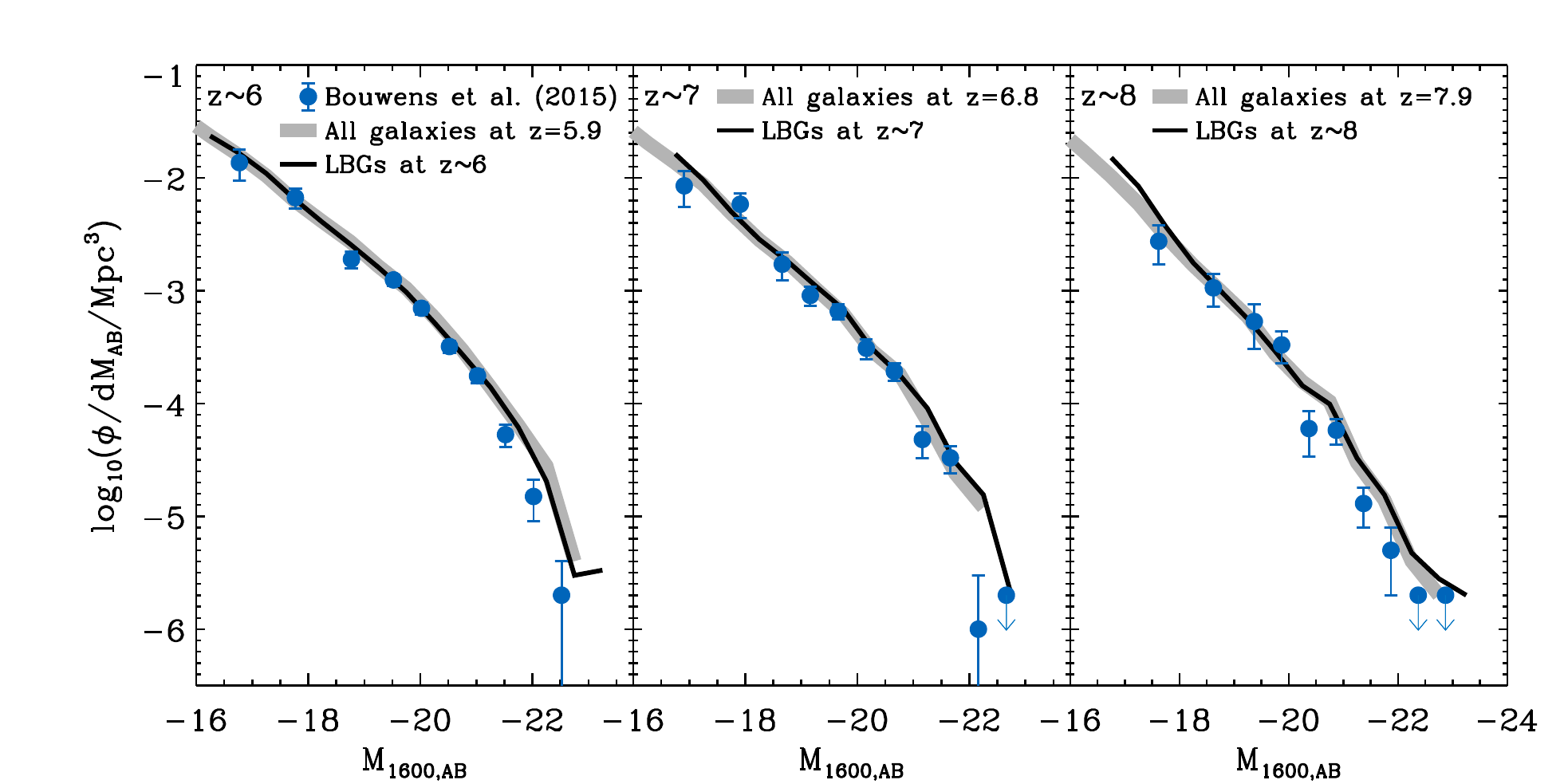}
\end{center}
\vspace{-3mm}
\caption{
The predicted rest-frame UV luminosity functions with the observed luminosity functions from \protect\cite{Bouwens2015}. Thick grey lines represent the predicted luminosity function using all galaxies in the snapshot at $z=5.9$, 6.8 and 7.9, respectively, which are mean redshifts for LBGs in $z\sim6$, $z\sim7$ and $z\sim8$ samples from \protect\cite{Bouwens2015}. Solid lines represent the predicted luminosity functions of selected LBGs in the redshift ranges $5.0 \lesssim z \lesssim 7.5$, $5.5 \lesssim z \lesssim 8.5$ and $6.0 \lesssim z \lesssim 9.5$, respectively. 
}
\label{LF}
\end{figure*}
%---------------------------------------------------------------------------------------------------------------------------------------------------%
Before using the colour selection criteria, we consider two photometric conditions to mimic the observations.
First, we take into account the photometric error for a predicted magnitude, which may affect the clustering signal for faint galaxies \citep{Park2016}. We obtain the apparent magnitude including the photometric error using
%======= Equation: Magnitude with error
\begin{equation}\label{eq:mag}
m' = -2.5{\rm log_{10}}(10^{-0.4\times m} + {\rm noise}),
\end{equation}
%==============================
where $m$ is an intrinsic apparent magnitude predicted from {\scshape Meraxes} and the noise denotes a random Gaussian flux uncertainty with a mean value of zero. We obtain a $1\,\sigma$ noise magnitude from the $5\,\sigma$ flux limits \citep{Bouwens2015} listed in Table\,\ref{Table:Flux_limit}, using ${\rm noise_{1\sigma}} = 10^{(-0.4\times m_{5\sigma})/5}$, where $m_{5\sigma}$ indicates the $5\,\sigma$ flux limit.

Second, we take into account the observational flux limits from \cite{Bouwens2015}. \cite{Bouwens2015} observed LBGs in combined survey fields consisting of the Hubble eXtreme Deep Field (XDF) and CANDELS survey. We apply the colour selection criteria to model galaxies brighter than the $5\,\sigma$ flux limits and use the different flux limits for the individual survey fields listed in Table\,\ref{Table:Flux_limit}. In cases of non-detection in the drop-out band ($i_{775}$ for $z\sim6$ and $z_{850}$ for $z\sim7$), \cite{Bouwens2015} set the magnitudes to be equal to the $1\,\sigma$ flux limit to measure a colour. We use the same substitution to model galaxies that are detected redward of the Lyman-break.

Fig.\,\ref{fig:CCD} shows the colours of model galaxies with the colour selection regions for LBGs at $z\sim6$, 7 and 8, respectively. We plot the predicted colours of galaxies at mean redshifts $z=5.9$, 6.8 and 7.9 (for LBGs at $z\sim6$, 7 and 8) estimated from observations \citep{Bouwens2015}. We find that the predicted colour distributions of model galaxies are broadened by photometric errors compared with the distribution of intrinsic colours. This is because the noise term in equation\,(\ref{eq:mag}) causes colours to be scattered, especially for faint galaxies, which are mainly hosted by low mass haloes.  

The redshift distribution can be written as%is defined as the number of selected galaxies per solid angle per redshift interval, 
%======= Equation: Magnitude with error
\begin{equation}\label{eq:Nz}
N(z)=n(z)\frac{{\rm d}^2V}{{\rm d}z\,{\rm d}\Omega},%\frac{{\rm d}^2N}{{\rm d}z\,{\rm d}\Omega}. 
\end{equation}
%===================================
where n(z) is the comoving number density of galaxies and ${\rm d}^2V/{\rm d}z\,{\rm d}\Omega$ is the comoving volume per solid angle per redshift interval.
Observationally, the redshift distribution is measured using an estimate of completeness. The completeness is defined as the ratio of selected galaxies to all galaxies at a given magnitude and redshift, and is estimated using the probability of recovering artificial LBGs \citep[e.g.][]{Yoshida2006,Bouwens2015}. Similarly, we compute the redshift distribution using the ratio of model galaxies selected as LBGs to all model galaxies at a given magnitude bin in each snapshot. Fig.\,\ref{fig:Nz6}  shows the predicted redshift distribution of model LBGs corresponding to individual survey fields at $z\sim6$, 7 and 8 respectively. We find that the predicted redshift distribution using the ratio between model LBGs and all model galaxies shows better agreement with the measured distribution \citep{Bouwens2015} than results based on equation\,(\ref{eq:Nz}). However, we note that the difference in clustering amplitude caused by the different determinations of $N(z)$ is between a few and 10 per cent. Relative to the observational uncertainties, this difference is not significant, but it may be important to include when modelling future observations.

Overall, the colour selection criteria of \cite{Bouwens2015} successfully exclude model galaxies outside the target redshift. We note that the colour selection criteria are designed to exclude low redshift interlopers as well as to select galaxies in the specific redshift range. Therefore, we demonstrate only that the selection criteria effectively select model galaxies in intended redshift ranges.

%%%%%%%%%%%%%%%%%%%%%%%%%%%%%%%%%%%%%%%%%%%%%%%%%%%%%%%%%%
%     Luminosity function
%%%%%%%%%%%%%%%%%%%%%%%%%%%%%%%%%%%%%%%%%%%%%%%%%%%%%%%%%
\subsection{Luminosity function}\label{sec:luminosity_func}
\cite{Liu2016} showed that the {\scshape Meraxes} model successfully predicts the observed rest-frame UV luminosity function in the redshift range $z=5-10$. To evaluate the influence of flux and redshift uncertainty on model LBG selection, we predict the rest-frame UV luminosity functions at $z\sim6$, 7 and 8 using all selected model LBGs over the redshift distributions shown in Fig.\,\ref{fig:Nz6}. The observed luminosity function is measured using an effective volume taking into account the completeness \citep[e.g.][]{Yoshida2006,Bouwens2015}. To mimic observations we define the effective volume in each UV magnitude bin as \citep{Park2016}
%============ Equation: Effective volume in simulation
\begin{equation}\label{eq:V_eff-sim}
V_{\rm eff} = \sum_{i}^{N}V_{\rm sim}\, p(m,z_i),
\end{equation}
%========================================
where $N$ is the snapshot number at which model LBGs are selected, $V_{\rm sim}$ is the simulation volume and $p(m,z_i)$ is a ratio of the number of selected LBGs to the total number of galaxies in a magnitude bin at snapshot $z_i$. 

Figure\,\ref{LF} shows the predicted rest-frame UV luminosity function together with the observed luminosity function from \cite{Bouwens2015}. For predicted luminosity functions at $z\sim6$, 7 and 8, the number of selected snapshots are 37, 40 and 40, and their redshift spans are $5.00 \leq z \leq 6.91$, $5.50 \leq z \leq 8.28$ and $6.00 \leq z \leq 9.61$, respectively. We find that the predicted luminosity functions using LBGs selected over the full photometric redshift distributions are in good agreement with the predicted luminosity functions using all galaxies at the target redshifts. This implies that the predicted luminosity function at each target redshift is representative of the predicted luminosity function over the redshift distribution \citep[see also][]{Park2016}. Note that we use snapshots at $z=5.9$, 6.8 and 7.9 as target redshifts for LBGs at $z\sim6$ and $z\sim7$, which are the estimated mean redshifts for LBG samples at $z\sim6$, 7 and 8 from \cite{Bouwens2015}. Overall, the predicted luminosity functions at $z\sim 6$ and 7 are consistent with the measured luminosity functions from \cite{Bouwens2015}. 

%%%%%%%%%%%%%%%%%%%%%%%%%%%%%%%%%%%%%%%%%%%%%%%%%%%%%%%
%      Modelling the ACF
%%%%%%%%%%%%%%%%%%%%%%%%%%%%%%%%%%%%%%%%%%%%%%%%%%%%%%%
\subsection{Modelling the angular correlation function} \label{sec:ACF}
The angular correlation function (ACF) provides the two-dimensional correlation of galaxies which are projected along line of sight. Limber's equation \citep{Limber1954} describes an integral relation between  the real-space correlation function and the ACF. If we assume that the mean number density does not rapidly vary with redshift and that the small angle approximation, this equation is a good approximation of the ACF. To compute the angular correlation function using snapshots generated at discrete redshifts, we use Limber's equation \citep{Limber1954}, 
%======= Equation: Limber
\begin{equation}\label{eq:Limber}
w(\theta)=\frac{2\int_{0}^{\infty}\left[ N(z)\right]^{2}/R_{\rm H}(z) \left( \int_{0}^{2r} {\rm d}u \ \xi(r_{12},z) \right) {\rm d}z} {\left[ \int_{0}^{\infty} N(z) {\rm d}z\right]^2},
\end{equation}
%==============================
where $N(z)$ is the redshift distribution of galaxies, $R_{\rm H}(z)$ is the Hubble radius and $\xi(r_{12},z)$ is the two-point correlation function. Using the small angle approximation, we denote $r_{12}=\sqrt{u^2 + r^2\theta^2}$, where $u=r_1-r_2$ and $r=(r_1+r_2)/2$ for comoving distances $r_1$ and $r_2$ to a pair of galaxies.

%---------------------------------------------------------------------------------------------------------------------------------------------------%
% Figure : Angular correlation function
%---------------------------------------------------------------------------------------------------------------------------------------------------%
\begin{figure*}
%\begin{figure*}
\begin{center}
\includegraphics[trim = 0cm 1cm 0cm 1cm, scale=0.915]{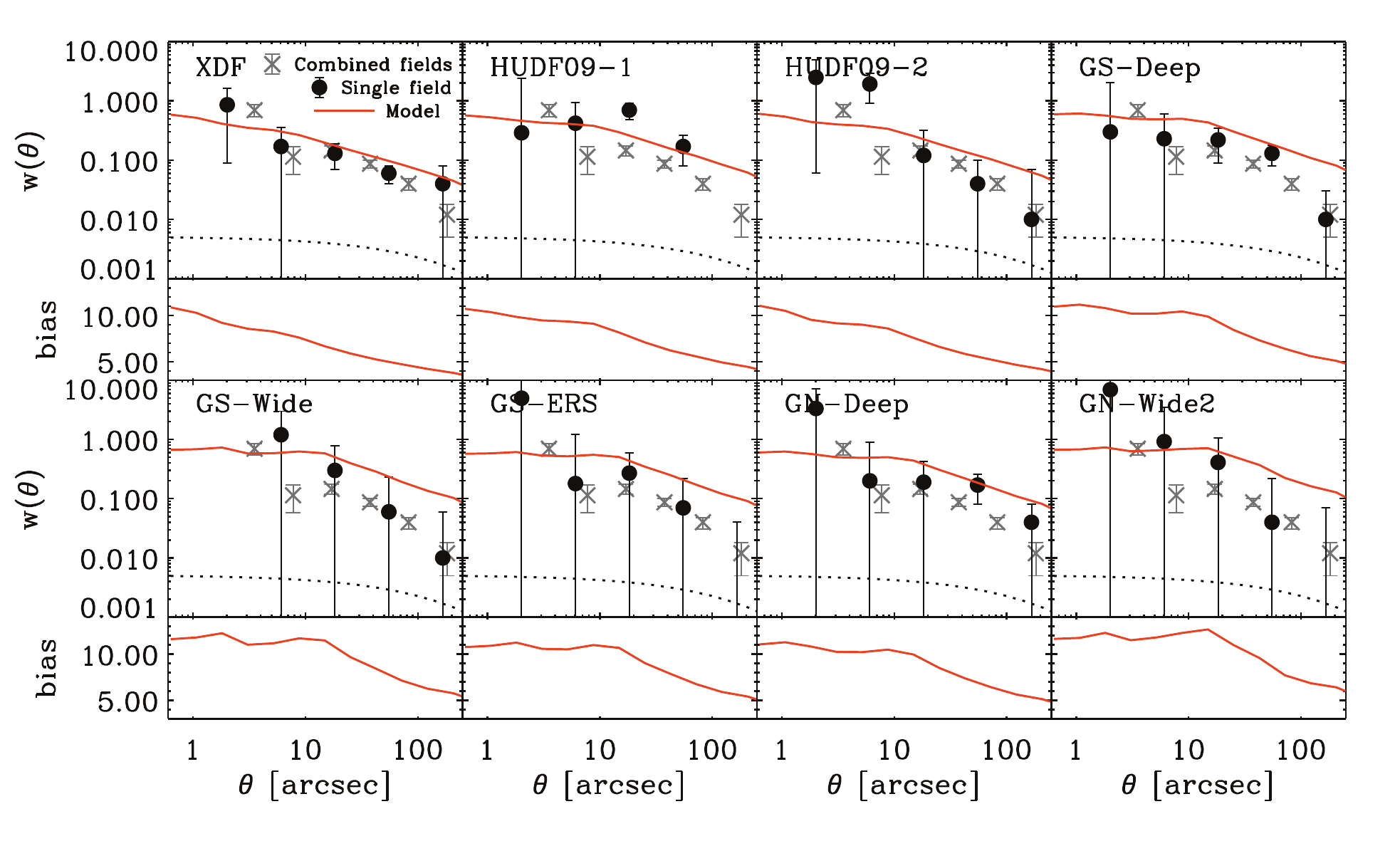}
\end{center}
%\vspace{-3mm}
\caption{
The predicted angular clustering of LBGs at $z\sim6$ (solid line). The dotted lines represent the predicted angular clustering of dark matter computed using the linear dark matter power spectrum. The name of the field is labelled on each panel. Filled circles with error bars show the observed ACF measured from the individual field. The crosses with error bars show the observed combined ACF and are reproduced in each panel for reference. All errors are $1\,\sigma$ and estimated using bootstrap resampling \protect\citep{Ling1986}. Note that the data points of the combined ACF and GS-Wide field at the smallest angular separation bins ($\theta \lesssim 3\,{\rm arcsec}$ and $\theta \lesssim 5\,{\rm arcsec}$, respectively) are not seen in this plot because the clustering amplitude is negative. The data point of HUDF09-1 at the largest angular separation bin ($\theta \gtrsim 80\,{\rm arcsec}$) is not seen for the same reason. We interpret these negative clustering amplitudes as being caused by insufficient numbers of galaxy pairs at small angular separations, and note that the ACFs binned on a linear scale do not show negative clustering at small scale \protect\citep[see][]{Rob2014}. In each case the bottom sub-panels show the predicted galaxy bias, defined as $b^2(\theta)=w_{\rm LBGs}(\theta)/w_{\rm DM}(\theta)$.
}
\label{fig:ACFz6}
\end{figure*}
%---------------------------------------------------------------------------------------------------------------------------------------------------%
Before computing Limber's equation, we firstly compute the two-point correlation function in each snapshot involved in the redshift distribution. The two-point correlation function is calculated using the excess probability of finding a pair of galaxies at separation $r$ to $r+\delta r$ compared to a random distribution,
%======= Equation: xi(r)
\begin{equation}\label{xir}
1+\xi(r)=\frac{DD}{\bar{n}^2}\frac{1}{V{\rm d}V},
\end{equation}
%==============================
where $DD$ is the number of galaxy pairs, $\bar{n}$ is the mean galaxy number density, $V$ is the simulation volume and ${\rm d}V$ is the differential volume between $r$ and $r+\delta r$. When integrating the two-point correlation function in equation\,(\ref{eq:Limber}) at scales beyond which the model cannot predict due to the finite simulation volume, we use a scaled dark matter two-point correlation function. On these scales the two-point correlation function is $\xi(r,z)=b(z)^2\,\xi_{\rm DM}(r,z)$, where $b$ is the linear galaxy bias and $\xi_{\rm DM}(r,z)$ is calculated by the linear initial dark matter power spectrum. To find the linear galaxy bias we use a mean bias over the range $5\,{\rm Mpc} \leq r \leq 10\,{\rm Mpc}$ \citep[see e.g.][]{Orsi2008}. In this study, we use the predicted two-point correlation function from simulations on scales up to 10 comoving Mpc, corresponding to $\sim 230$ arcsec at $z\sim7$. This length scale is sufficiently large to compare the model predictions with current observations.

%---------------------------------------------------------------------------------------------------------------------------------------------------%
% Figure : Angular correlation function at z~7
%---------------------------------------------------------------------------------------------------------------------------------------------------%
%\begin{figure*} 
\begin{figure*}
\begin{center}
\includegraphics[trim = 0cm 0.5cm 0cm 1cm, scale=0.915]{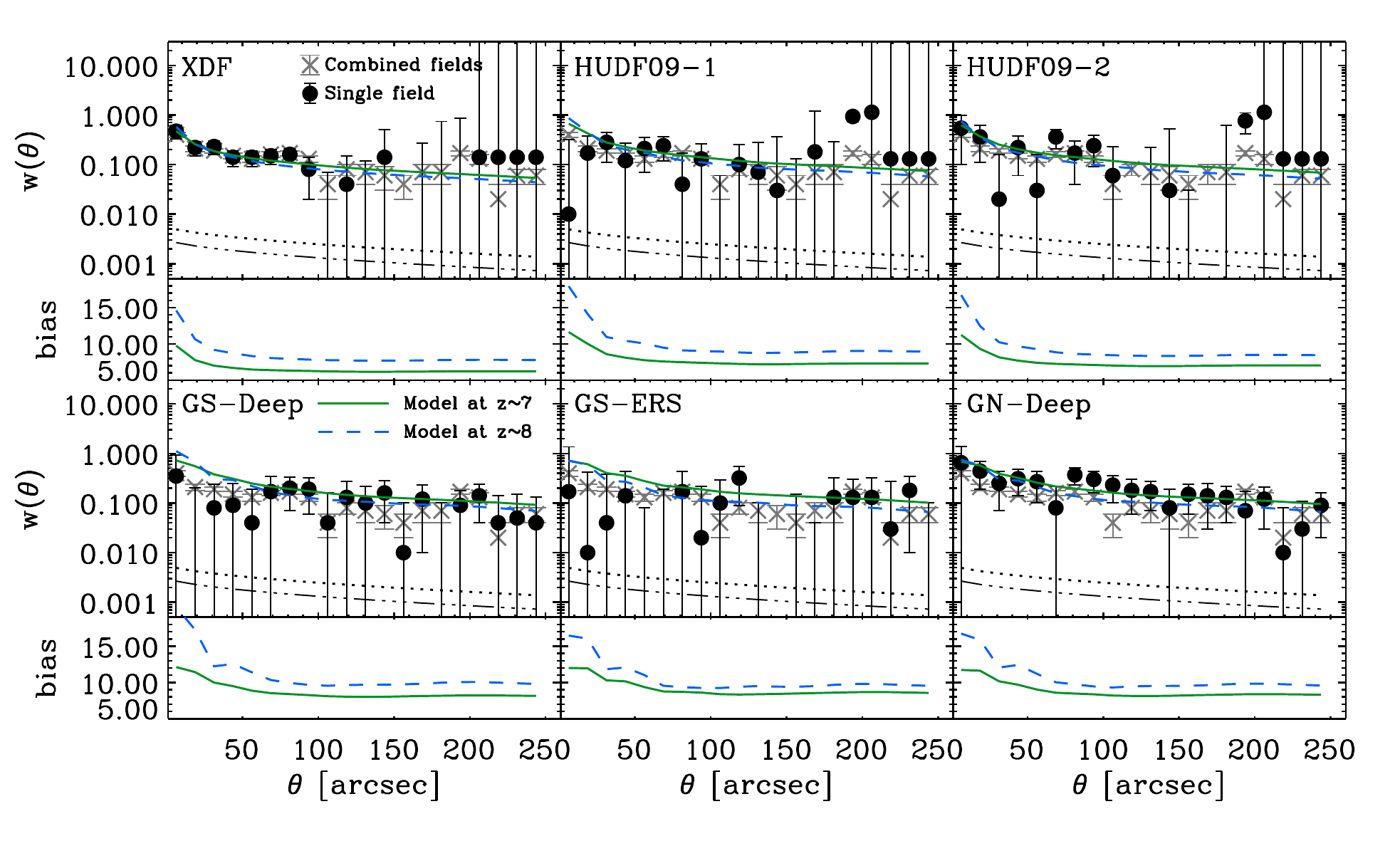}
\end{center}
\vspace{-3mm}
\caption{
The same as Fig.~\ref{fig:ACFz6}, but at $z\sim7$ and 8. Dotted and dashed-dot lines represent the predicted ACF of dark matter at $z\sim7$ and 8, respectively. Note that we plot the ACF on a linear scale, because of the sparse number of observed pairs at small angular separations.
}
\label{fig:ACFz7}
\end{figure*}
%---------------------------------------------------------------------------------------------------------------------------------------------------%
%%%%%%%%%%%%%%%%%%%%%%%%%%%%%%%%%%%%%%%%%%%%%%%%%%%%%%%
%
%      Comparison with observations                                                                                                                    %
%
%%%%%%%%%%%%%%%%%%%%%%%%%%%%%%%%%%%%%%%%%%%%%%%%%%%%%%%

\section{Comparison with observations}\label{sec:comparison}
In this section we show the predicted angular clustering and galaxy bias, and compare the resulting clustering properties with observations.
%%%%%%%%%%%%%%%%%%%%%%%%%%%%%%%%%%%%%%%%%%%%%%%%%%%%%%%
%      Angular correlation function                                                                                                                         %
%%%%%%%%%%%%%%%%%%%%%%%%%%%%%%%%%%%%%%%%%%%%%%%%%%%%%%%
\subsection{Angular correlation function}\label{comparison_ACF}

\cite{Rob2014} measured the ACF of LBGs at $z\sim 4-7.2$ using the samples of \cite{Bouwens2015}. They measured ACFs in eight and six individual survey fields for LBGs at $z\sim6$ and $z\sim7.2$, respectively. For the ACFs at $z\sim 7.2$, \citeauthor{Rob2014} combined the $z\sim7$ and  $z\sim8$ samples together to measure the ACFs. The redshift of 7.2 is the mean redshift estimated from the redshift distribution. In this paper, we do not reproduce the ACFs at $z\sim 7.2$, using combined samples at $z\sim 7$ and $8$. Instead, we predict the ACF at $z\sim7$ and 8 separately and investigate the clustering properties at each redshift. \cite{Rob2014} also calculated a combined ACF from the ACFs measured using the individual survey fields (see \citealp{Rob2014} for more detail). We do not reproduce the combined ACF in this study, but we show the combined measurement as a reference.

%Fig.\,\ref{fig:ACFz6} shows the predicted ACF of LBGs at $z\sim6$ selected from the model together with the measured ACFs and the combined ACF from observations. We show the ACFs on a log scale to accurately depict the clustering on large scales. We find that there are noticeable differences between the measured ACFs from each survey field. These field-to-field variation can be explained by cosmic variance. In addition, the observed sample variance in each field contributes to the field-to-field variation, because there are order of magnitude differences in the number of LBGs used in the ACF analysis (as listed in Table\,\ref{Table:Flux_limit}). On the other hand, the predicted ACFs do not show significant field-to-field variation. When considering the estimated uncertainties from observational measurements, the predicted ACFs for each individual field are in good agreement with the measured ACFs from observations. Fig.\,\ref{fig:ACFz7} shows the predicted angular clustering of LBGs at $z\sim7$ and 8 selected from the model together with the measured ACFs at $z\sim7.2$ and the combined ACF from observations. For this case, we plot the ACF on a linear scale, because of the sparse number of observed pairs at small angular separations. Predicted ACFs for each individual field are again in good agreement with the measured ACFs from observations. 

Fig.\,\ref{fig:ACFz6} shows the predicted ACF of LBGs at $z\sim6$ selected from the model together with the measured ACFs and the combined ACF from observations. We show the ACFs on a logarithmic scale to accurately depict the clustering on large scales. The predicted ACFs for each individual field are consistent to within $\sim2.5\,\sigma$ of the observed ACFs except GS-Deep field on large scale. On large scales it appears as if the model may over predicts the amplitude by up to an order of magnitude. However, we find that there are noticeable differences between the measured ACFs from each survey field. These field-to-field variation can be explained by cosmic variance. In addition, the observed sample variance in each field contributes to the field-to-field variation, because there are order of magnitude differences in the number of LBGs used in the ACF analysis (as listed in Table\,\ref{Table:Flux_limit}). Because of this field-to-field variation, the measured ACFs do not clearly show the expected dependence on luminosity (as we will discuss in the next section). Fig. 1 of \cite{Rob2014} also shows that the combined ACF from all fields shows a large variation in the amplitude on large scales. The presence of this variation reinforces our interpretation of it being driven by sample variance. On the other hand, the predicted ACFs do not show significant field-to-field variation. Overall, when considering the estimated uncertainties from observational measurements, the predicted ACFs for each individual field are in good agreement with the measured ACFs from observations. We note that the predicted ACF in GS-Deep field is also consistent with the observation to within $2\,\sigma$ when comparing the ACF with the measured power-law ACF, as discussed in the next section.  

Fig.\,\ref{fig:ACFz7} shows the predicted angular clustering of LBGs at $z\sim7$ and 8 selected from the model together with the measured ACFs at $z\sim7.2$ and the combined ACF from observations. For this case, we plot the ACF on a linear scale, because of the sparse number of observed pairs at small angular separations. Predicted ACFs for each individual field are again in good agreement with the measured ACFs from observations. 

%%%%%%%%%%%%%%%%%%%%%%%%%%%%%%%%%%%%%%%%%%%%%%%%%%%%%%%
%      Dependence of clustering on luminosity                                                                                                      %
%%%%%%%%%%%%%%%%%%%%%%%%%%%%%%%%%%%%%%%%%%%%%%%%%%%%%%%
\subsection{Dependence of clustering on luminosity}\label{sec:dependence_on_luminosity}
%---------------------------------------------------------------------------------------------------------------------------------------------------%
% Figure : Aw
%---------------------------------------------------------------------------------------------------------------------------------------------------%
\begin{figure*}
\begin{center}
\includegraphics[width=17.5cm]{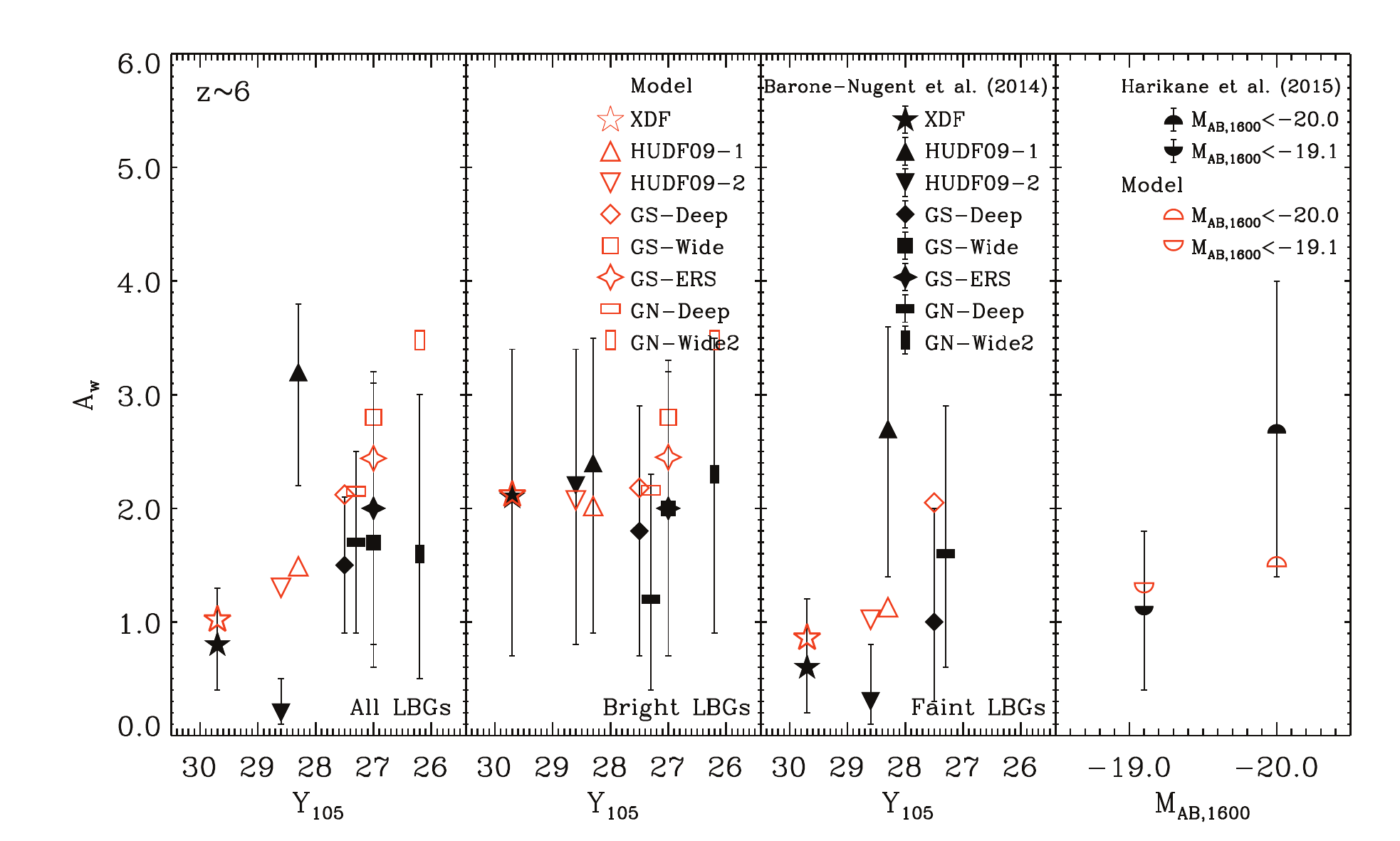}
\includegraphics[width=17.5cm]{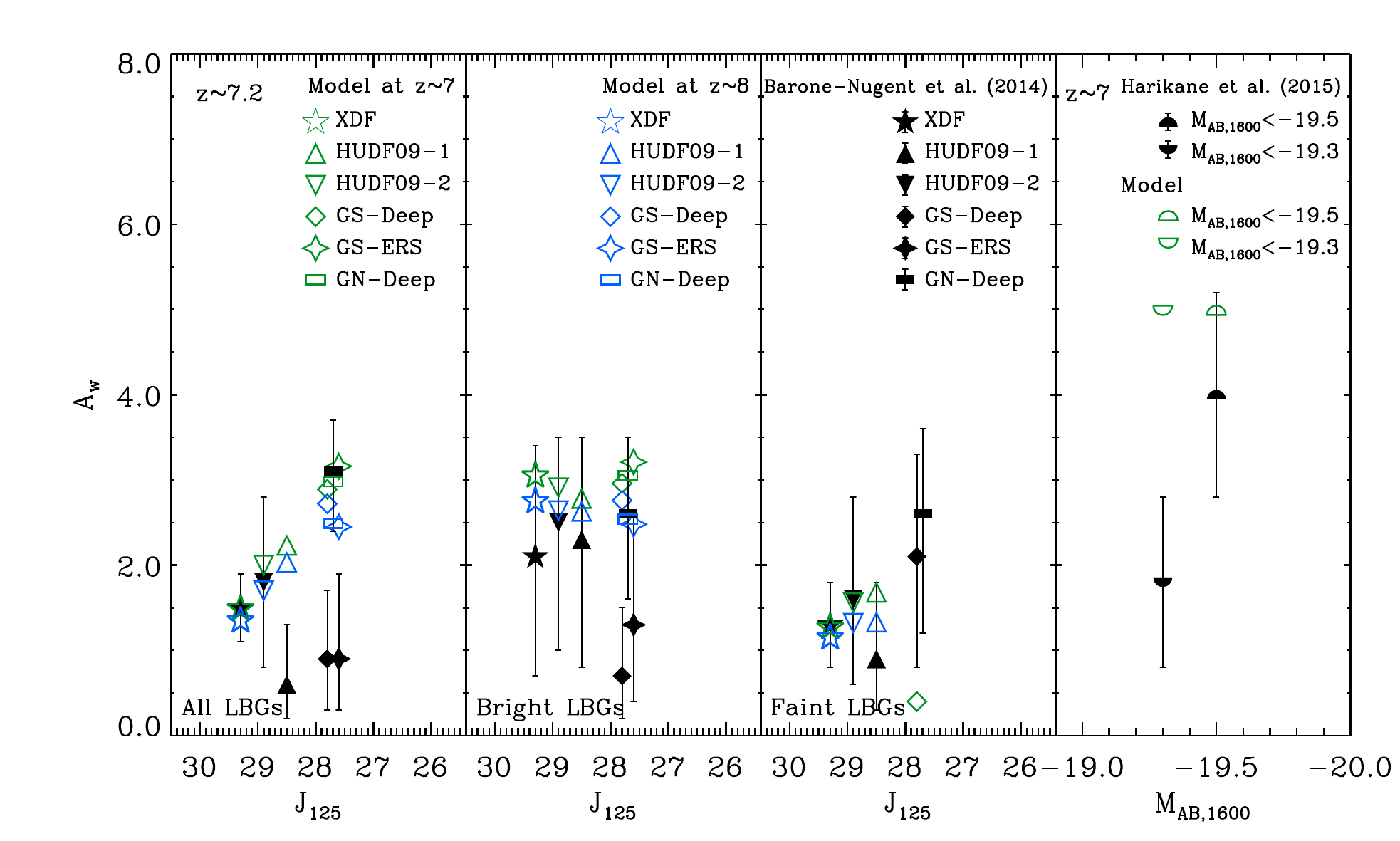}
\end{center}
\vspace{-3mm}
\caption{Top: Comparison of the angular correlation amplitude, $A_w$, with observations at $z\sim6$. The filled symbols with error bars represent the measured $A_w$ by \protect\cite{Rob2014} and the positions of x-axis indicate the detection limit for $Y_{105}$ band magnitude corresponding to $M_{\rm AB,1600}$. The empty symbols represent the model predictions. Left, middle-left and middle-right panels show the galaxy bias of all, bright and faint LBGs, respectively. The right panel shows the $A_w$ measured by \protect\cite{Harikane2015} and the model prediction. Bottom: The same as top panels, but we compare the predicted $A_w$ values at $z\sim7$ and 8 with the measured $A_w$ at $z\sim7.2$ by \protect\cite{Rob2014}. Note that for $z\sim8$ LBGs $H_{160}$ band corresponds to the rest-frame UV magnitude, but we plot the predicted $A_w$ at $z\sim8$ as a function of $J_{125}$ band for simplicity, since the detection limit of $H_{160}$ band decreases the same rate as the $J_{125}$ band (see Table\,\ref{Table:Flux_limit}). 
}
\label{fig:Best-fit_Aw}
\end{figure*}
%---------------------------------------------------------------------------------------------------------------------------------------------------%

In analytic and numerical models for dark matter haloes, more massive haloes cluster more strongly than less massive haloes \citep[e.g.,][]{MW96}. Over the past decade observational results show that the clustering strength of galaxies depends on luminosity both at high redshifts \citep{Ouchi2004,Ouchi2005,Cooray2006,Lee2006,Kashikawa2006,Hildebrandt2009} and in the local Universe \citep[e.g.][]{Norberg2001,Zehavi2002}. This clustering dependence on luminosity supports the expectation that more massive haloes host brighter galaxies \citep[e.g.,][]{G&D2001}. Recent studies \citep{Rob2014,Harikane2015} also showed this clustering dependence on luminosity for LBGs up to $z\sim7$.

The ACFs for each individual survey field allow us to investigate dependence of clustering on luminosity, since the flux limits of these fields listed in Table\,\ref{Table:Flux_limit} are all different. The predicted clustering amplitude for each individual field  at $z\sim6$ (Fig.\,\ref{fig:ACFz6}) increases from the deepest field (XDF) to the shallowest field (GN-Wide2). We further find that the model predictions are comparable with these measurements. The predicted clustering amplitudes at $z\sim7$ and 8 (Fig.\,\ref{fig:ACFz7}) also show a similar trend. 

On large scales, the ACFs can be approximated by a power-law using the angular correlation amplitude, $A_w$, and the correlation slope, $\beta$, i.e. $w(\theta)=A_{w}\,\theta^{-\beta}$. To more clearly show the dependence of clustering on luminosity, we calculate the best fitting angular correlation amplitude, $A_w$, using the least square method for each survey field and compare this predicted value with observational measurements \citep{Rob2014,Harikane2015}. Note that \cite{Rob2014} and \cite{Harikane2015} fixed $\beta=0.6$ and 0.8, respectively, to correct the clustering amplitude measured from their sample \citep[e.g.][]{Roche&Eales1999, Lee2006}. We will present the best-fit value of $A_w$ predicted from the model with fixed $\beta$ ($\beta$=0.6 and 0.8) for comparison, and analyse the values of the best-fit value of $A_w$ when allowing the value of $\beta$ to vary. We also note that we find the best fitting values by considering only angular separations larger than 10 arcsec We do not show uncertainties for these best-fit values because statistical errors, e.g. bootstrap resampling, are too small to be presented.

The left panels of Fig.\,\ref{fig:Best-fit_Aw} show the measured and predicted $A_w$ values for each individual survey field at $z\sim6$ (top), and $z\sim7$ and 8 (bottom) as a function of the magnitude limit of $Y_{105}$ and $J_{125}$ band, respectively, corresponding to the rest-frame UV magnitude. In addition to computing clustering in individual field galaxy samples, \cite{Rob2014} also measured ACFs of bright and faint subsamples split using a UV luminosity of $M_{\rm AB(1600)}=-19.4$ which is the median magnitude for the overall sample. The corresponding values for $A_w$ are shown in the middle-left and -right panels of Fig.\,\ref{fig:Best-fit_Aw}. We split model LBGs into bright and faint subsets using the same magnitude cut, and show the corresponding model clustering predictions for bright and faint subsets. Note that since the median UV luminosity is for all samples, $A_w$ values for bright LBGs are not measured in GS-ERS, GN-Deep and GN-Wide2 fields from observations, and also are not predicted in GS-Wide, GS-ERS, GN-Deep and GN-Wide2 fields from the model. Fig.\,\ref{fig:Best-fit_Aw} shows that whilst there is significant field-to-field variation due to cosmic variance \citep{Trenti2008}, the data shows the trend that brighter LBGs are more highly clustered, based both on the dependence with limiting magnitude, and comparison between bright and faint samples.

Before comparing model predictions with observations, we firstly consider the model predictions in isolation. At $z\sim6$ (top-left panel in Fig.\,\ref{fig:Best-fit_Aw}) the predicted angular correlation amplitude increases from the deepest survey field (XDF) toward the shallowest survey field (GN-Wide2). In addition, the comparison between the bright and faint subsets (middle-left and -right in top panels) show a similar trend. The predicted amplitudes for bright subsets are higher than those predicted for faint subsets. At $z\sim7$ and 8 (bottom panels) the model predictions show similar trends to the predictions at $z\sim6$. We note that, in wide and shallow fields (GS and GN fields), the correlation amplitudes of bright LBGs are almost identical to those of all LBGs. This is because the wide and shallow fields contain mainly bright LBGs, so that the two samples are very similar. 

%---------------------------------------------------------------------------------------------------------------------------------------------------%
% Figure : Aw
%---------------------------------------------------------------------------------------------------------------------------------------------------%
\begin{figure}
\begin{center}
\includegraphics[width=8.7cm]{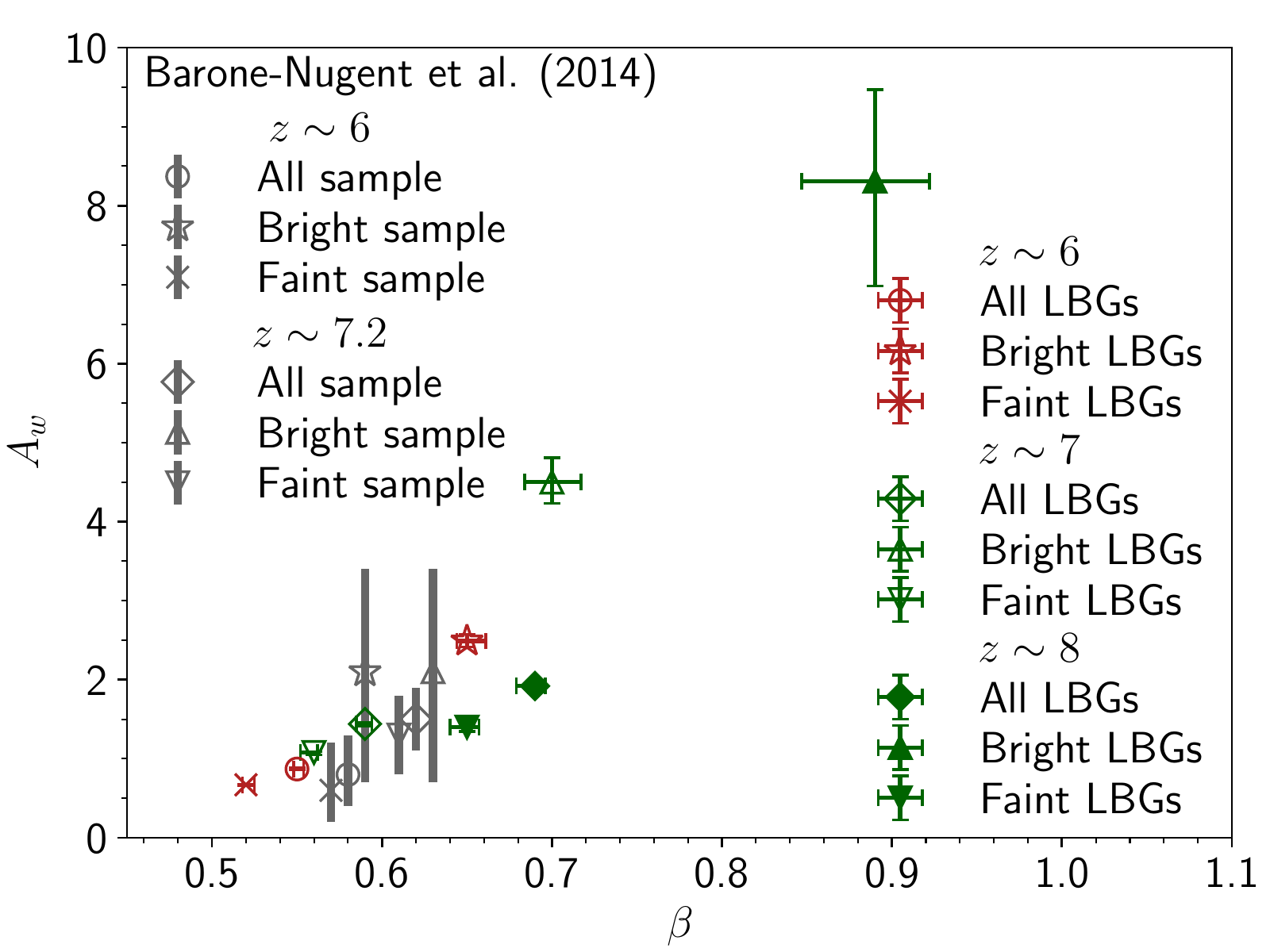}
\end{center}
\vspace{-3mm}
\caption{Predicted best-fit values of $A_w$ and $\beta$ at $z\sim6$\,-\,8 for XDF field. Different symbols correspond to all, bright and faint subset of model LBGs at different redshifts as indicated by the legend. Estimated $A_w$ values from \protect\cite{Rob2014} are shown for comparison. We note that the observational $A_w$ values are estimated with an assumption of $\beta = 0.6$. These values are slightly shifted on $\beta$ for clarity. All errors are $1\,\sigma$ and estimated using bootstrap resampling.
}
\label{fig:Aw_vs_beta}
\end{figure}
%---------------------------------------------------------------------------------------------------------------------------------------------------%
When comparing the model predictions with observations, we find that the trends of clustering with luminosity are similar between models and observation. Thus, clustering measurements imply that brighter LBGs reside in more massive haloes at $z>6$, as expected from simulations \citep{Liu2016}. We note that the model predictions for the HUDF deviates from the measured values, although it is still consistent within $3\,\sigma$. This is likely due to large sample variance, because the HUDF09-1 and HUDF09-2 have only a few tens of LBGs. Overall, we find that the trends are much clearer than in the observed samples because our single simulation realisation does not include cosmic variance. 

We can also study the dependence of clustering on luminosity using the measured ACFs by \cite{Harikane2015}. They split the observed sample at $z\sim6$ into two subsamples with thresholds of $M_{\rm AB(1600)}<-20.0$ and $M_{\rm AB(1600)}<-19.1$, respectively. They also split the sample at $z\sim7$ into two subsamples with thresholds of $M_{\rm AB(1600)}<-19.5$ and $M_{\rm AB(1600)}<-19.3$, respectively. 

We find that the predicted best-fitting $A_w$ values using the same rest-frame UV magnitude thresholds show a dependence of increased clustering with luminosity (right panels of Fig.\,\ref{fig:Best-fit_Aw}), supporting conclusions based on HUDF and COSMOS fields. We note that the predicted values of $A_w$ at $z\sim7$ are identical. This is caused by the fixed value of $\beta$ as discussed below. We checked that when we allow the value of $\beta$ to vary the predicted values of $A_w$ show the same trend with the observation and are consistent within 2\,$\sigma$ errors.

Fig.\,\ref{fig:Aw_vs_beta} shows the predicted best-fitting values of $A_w$ and $\beta$ for model LBGs at $z\sim 6$\,-\,8 when allowing the value of $\beta$ to vary. As shown in Fig.\,\ref{fig:Best-fit_Aw}, the best-fit values of $A_w$ increase with luminosity. We also find that the best-fit values of $\beta$ increase with luminosity. This is consistent with observational measurements at $z\sim4$\,-\,5 from \cite{Kashikawa2006} and a semi-analytical model prediction at $z\sim4$ from \cite{Park2016}. This trend shows that $A_w$ and $\beta$ increase with increasing redshift. When comparing predicted values with estimated $A_w$ at $z\sim6$ and $z\sim7.2$ from \cite{Rob2014}, the predicted best-fit values are still consistent with observations within $1\,\sigma$ errors. The predicted values of bright LBGs at $z\sim 7\,-\,8$ are consistent with observations within $\sim2.5\,\sigma$ errors. We note that Fig.\,\ref{fig:Best-fit_Aw} shows $A_w$ at $z\sim8$ to be higher than that at $z\sim7$ which is in contrast to this result. We interpret that this results from the fixed $\beta$. We checked that the best-fitting $A_w$ is underestimated when the best-fitting $\beta$ is larger than a fixed $\beta$ and vice versa. Although this difference arising from the fixed $\beta$ is not significant when taking into account the current observational uncertainty, this factor should be considered in future surveys.

Overall, the model prediction of the correlation amplitude reproduces the measured dependence of clustering on luminosity from observations within $2\,\sigma$ errors.

%%%%%%%%%%%%%%%%%%%%%%%%%%%%%%%%%%%%%%%%%%%%%%%%%%%%%%%
%      Bias                                                                                                                                        %
%%%%%%%%%%%%%%%%%%%%%%%%%%%%%%%%%%%%%%%%%%%%%%%%%%%%%%%
\subsection{Galaxy bias and halo mass}\label{sec:comparison_bias}
We show the predicted galaxy bias for each survey field at $z\sim6$\,-\,8 (bottom sub-panels in Fig.\,\ref{fig:ACFz6} and Fig.\,\ref{fig:ACFz7}), defined as the ratio of angular correlation function of model LBGs to the angular correlation function of dark matter, $b^2(\theta) = w(\theta)/w_{\rm DM}(\theta)$. We computed the ACF of dark matter using the initial linear dark matter power spectrum linearly extrapolated using the growth factor. We compute the ACF of dark matter using the same redshift distribution as the galaxies (Fig.\,\ref{fig:ACFz6} and Fig.\,\ref{fig:ACFz7})) for each survey field. 

In general, biases for deep fields (XDF and HUDFs) are lower than biases for shallow fields, i.e. brighter galaxies show higher bias. This result follows directly from and reinforces the clustering dependence on luminosity discussed in \S\,\ref{sec:dependence_on_luminosity}. Scale dependent bias is a general prediction of the hierarchical structure formation framework \citep[e.g.,][]{Colin1999}. We find that model biases increase with decreasing angular separation, but are almost constant on large angular separations ($\theta \gtrsim 70\,{\rm arcsec}$). In addition, this small scale increase is more significant for shallow fields consisting only of bright galaxies. Although it is not possible to compare this result directly with observations at $z\sim 6$, for observed LBGs at lower redshifts ($z\sim4$ and $z\sim5$) scale dependent bias has been reported \citep{Hamana2004, Kashikawa2006,Lee2006}. 

To compare model predictions with observations, we therefore use large scale bias. \cite{Rob2014} estimated the galaxy bias using the ratio of  the galaxy variance at $8\,h^{-1}{\rm Mpc}$, $\sigma_{8,{\rm g}}$, to  the linear matter fluctuation at $8\,h^{-1}{\rm Mpc}$, $\sigma_8(z)$ \citep[e.g.,][]{Lee2006}. The galaxy variance, $\sigma_{8,{\rm g}}$, is calculated from the ACF using \citep{Peebles1980}
%======= Equation: galaxy variance
\begin{equation}\label{eq:sig8g}
\sigma^{2}_{8,{\rm g}}=\frac{72(r_{0}/8\,h^{-1}{\rm Mpc})^{\gamma}}{(3-\gamma)(4-\gamma)(6-\gamma)\,2^{\gamma}},
\end{equation}
%==============================
where $r_{0}$ is the correlation length and $\gamma = \beta +1$ are parameters to approximate the real-space correlation function using a power-law, $\xi(r) = (r/r_0)^{\gamma}$. 

In the model, we calculate the best fitting parameters of $r_0$ and $\gamma$ from the predicted two-point correlation function in the range $1\,{\rm Mpc} < r < 10\,{\rm Mpc}$ for each snapshot, and determine a weighted average value of $r_0$ and $\gamma$ using the redshift distribution. We note that we do not fix $\gamma$ as was done with the observations. We checked that any resulting biases from fixed $\gamma$ are not significant.
%---------------------------------------------------------------------------------------------------------------------------------------------------%
% Figure : bias as a function of redshift
%---------------------------------------------------------------------------------------------------------------------------------------------------%
\begin{figure*}
\begin{center}
\includegraphics[width=17.5cm]{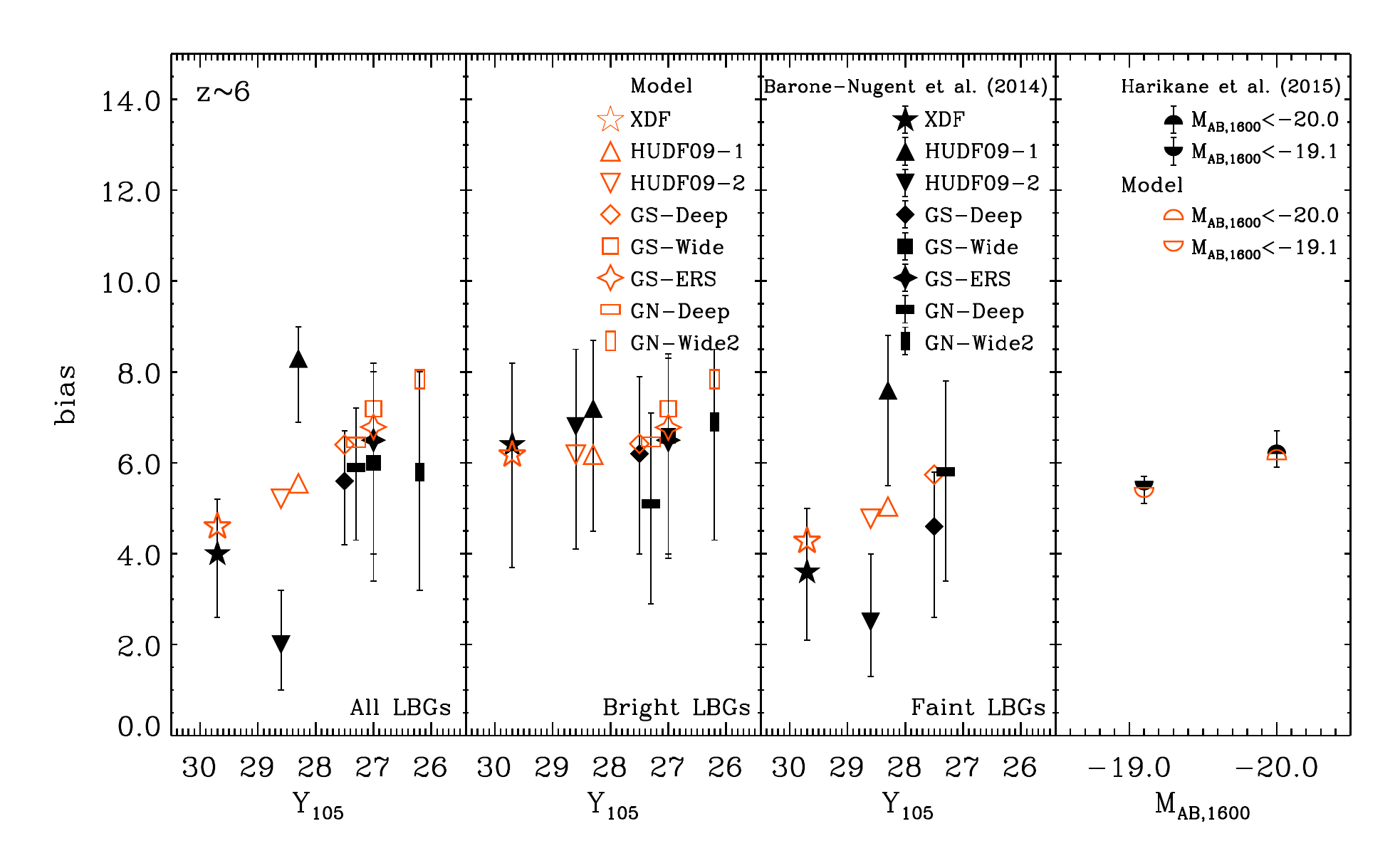}
\includegraphics[width=17.5cm]{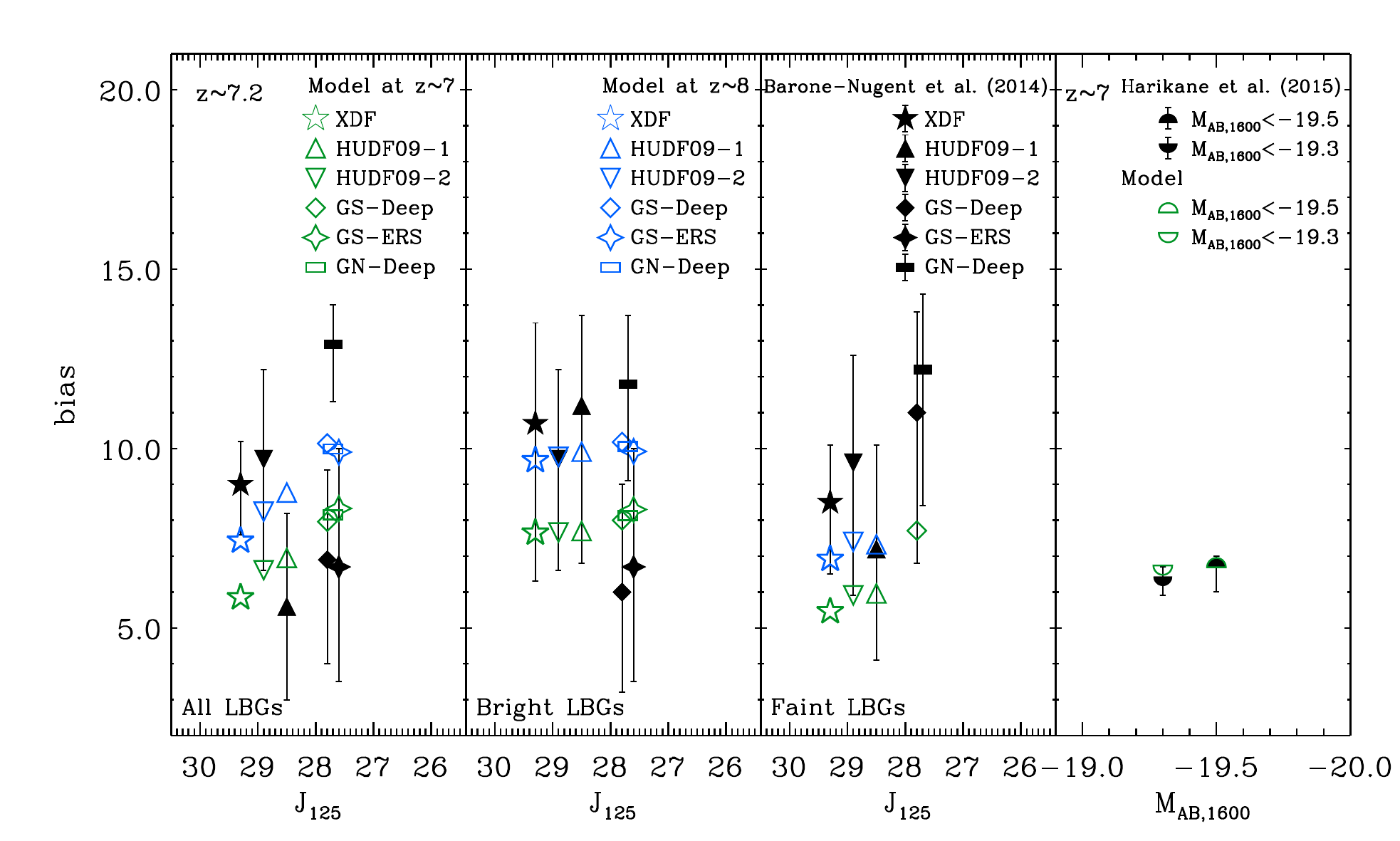}
\end{center}
\vspace{-3mm}
\caption{Top: Comparison of the predicted galaxy bias with observations at $z\sim6$. The galaxy bias is defined as $\sigma_{8,{\rm g}}/\sigma_{8}$, where $\sigma_{8,{\rm g}}$ is the galaxy variance at $8h^{-1}{\rm Mpc}$. The filled symbols with error bars represent the measured bias by  \protect\cite{Rob2014} and the positions on the x-axis indicate the detection limit for $Y_{105}$ band magnitude corresponding to $M_{\rm AB,1600}$. The empty symbols represent the model predictions. Left, middle-left and middle-right panels show the galaxy bias of all, bright and faint LBGs, respectively. The right panel shows the galaxy bias measured by \protect\cite{Harikane2015} and the model prediction. Bottom: The same as the top panels,  but we compare the predicted biases at $z\sim7$ and 8 with the estimated bias at $z\sim7.2$ from \protect\cite{Rob2014}. Note that for $z\sim8$ LBGs the $H_{160}$ band corresponds to the rest-frame UV magnitude, but we plot the predicted bias at $z\sim8$ as a function of $J_{125}$ band for simplicity since the detection limit of $H_{160}$ band decreases at the same rate as $J_{125}$ band (see Table\,\ref{Table:Flux_limit}).
}
\label{fig:bias1}
\end{figure*}
%---------------------------------------------------------------------------------------------------------------------------------------------------%
Fig.\,\ref{fig:bias1} shows the predicted large scale bias at $z\sim6$ (left three panels in top row), $z\sim7$ and $z\sim8$ (left three panels in bottom row), respectively. In agreement with the angular correlation amplitude (Fig.\,\ref{fig:Best-fit_Aw}), the predicted bias increases from the deepest field to the shallowest field. Bright subsamples for each survey field also show a higher bias than faint subsamples. When comparing the model prediction with observations, all predicted biases are consistent with observations (except HUDF fields). We interpret this to be because of large sample variance in these fields as discussed in \S\,\ref{sec:dependence_on_luminosity}.

\cite{Waters2016} studied galaxy clustering at $z=8$, 9 and 10 using their hydrodynamic simulation. They predicted the linear galaxy bias of $13.4\pm1.8$ at $z=8$ from the real-pace correlation function. This value is higher than the prediction from our model of $b=9.67$ for bright LBGs ($M_{\rm AB(1600)} < -19.4$) in the XDF field. However, considering their assumed detection limit ($M_{\rm AB(1600)} \lesssim -20.5$), our result is consistent with their prediction.

\cite{Harikane2015} estimated the effective galaxy bias from their Halo Occupation Distribution (HOD) modelling,
%======= Equation: galaxy variance
\begin{equation}\label{eq:effectiv_bias}
b^{\rm eff}_{\rm g}=\frac{1}{n_{\rm g}}\,\int{\rm d}M_{\rm h}\,\frac{{\rm d}n}{{\rm d}M_{\rm h}}(M_{\rm h},z)\,N(M_{\rm h})\,b_{\rm h}(M_{\rm h},z),
\end{equation}
%==============================
where $n_{\rm g}$ is the mean galaxy number density over the redshift distribution, $N(M_{\rm h})$ is the mean number of galaxies in a dark matter halo of mass $M_{\rm h}$, $\frac{{\rm d}n}{{\rm d}M_{\rm h}}(M_{\rm h},z)$ is the halo mass function, and $b_{\rm h}(M_{\rm h},z)$ is the halo bias. To compare the model prediction with observations we assume the halo bias of \cite{Tinker2010} following the assumption of \cite{Harikane2015}, but do not use HOD modelling since the model directly provides the number of galaxies as a function of halo mass. 

To investigate dependence with luminosity for this sample we split model LBGs into two subsets using the rest-frame UV magnitude thresholds corresponding to those in observations from \cite{Harikane2015}; $M_{\rm AB(1600)}<-20.0$ and $M_{\rm AB(1600)}<-19.1$ for LBGs at $z\sim6$, and $M_{\rm AB(1600)}<-19.5$ and $M_{\rm AB(1600)}<-19.3$ for LBGs at $z\sim 7$, respectively. As shown in the case of the large scale bias, the predicted effective bias increases with luminosity, and the resulting biases are in good agreement with the estimated bias from observations at both $z\sim6$ and $z\sim7$ (right panels in Fig.\,\ref{fig:bias1}). 

%---------------------------------------------------------------------------------------------------------------------------------------------------%
% Figure : bias as a function of redshift
%---------------------------------------------------------------------------------------------------------------------------------------------------%
\begin{figure*}
\begin{center}
\includegraphics[width=17cm]{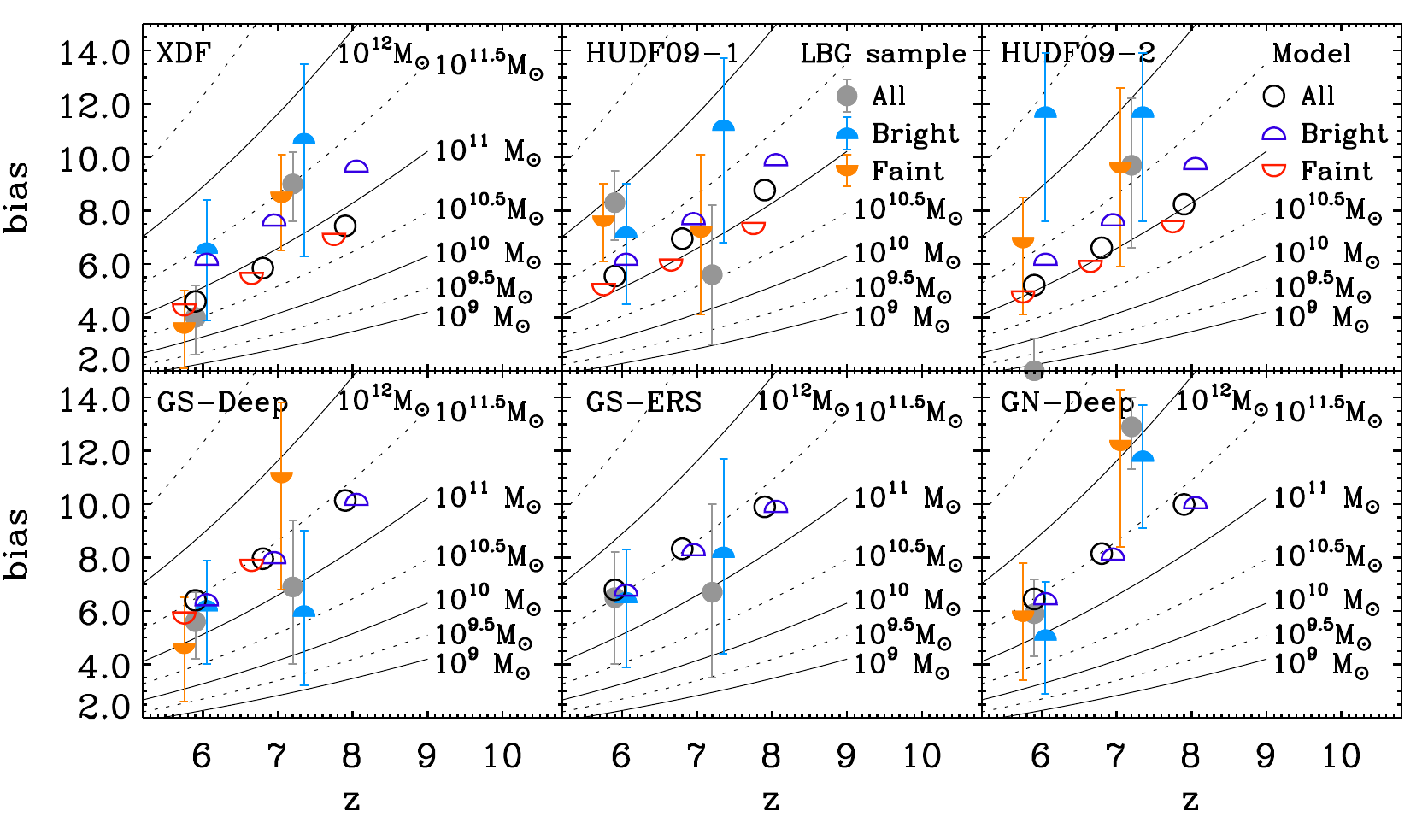}
\end{center}
\vspace{-3mm}
\caption{Galaxy bias as a function of redshift for each survey field. The galaxy bias is defined as $\sigma_{8,{\rm g}}/\sigma_{8}$. Filled circles, upper half circles and lower half circles with error bars are measured bias of all, bright and faint sample, respectively. Empty circles, upper half circles and lower half circles are the predicted bias for all, bright and faint subset from the model. The solid and dotted lines represent the average bias for haloes with $M\geq M_{\rm halo}$ from \protect\cite{Tinker2010}. Note that the model does not predict the bias of the faint subset for the GS-ERS and GN-Deep fields due to an insufficient number of galaxies. Similarly, we do not show the measured bias of the faint subsample from the GS-ERS field for the same reason.
}
\label{fig:bias_evolution}
\end{figure*}
%---------------------------------------------------------------------------------------------------------------------------------------------------%
Fig.\,\ref{fig:bias_evolution} shows the evolution of observed galaxy bias as a function of redshift for each survey field. The biases increase with increasing redshift at fixed mass. As already noted, the model prediction also shows that brighter LBGs reside in more massive dark matter haloes. The halo mass corresponds to the mass range $10^{10.5}M_{\rm \odot} \lesssim M_{\rm halo} \lesssim 10^{11.5}M_{\rm \odot}$, which is consistent with the predicted halo mass for LBGs \citep[see][]{Liu2016}. We note that this mass range is slightly higher than the estimated halo mass from \cite{Rob2014}, since we use a different bias calculation. The predicted biases are consistent with observations within approximately $2\,\sigma$ errors. 

%---------------------------------------------------------------------------------------------------------------------------------------------------%
% Figure : bias as a function of redshift
%---------------------------------------------------------------------------------------------------------------------------------------------------%
\begin{figure}
\begin{center}
\includegraphics[width=8.5cm]{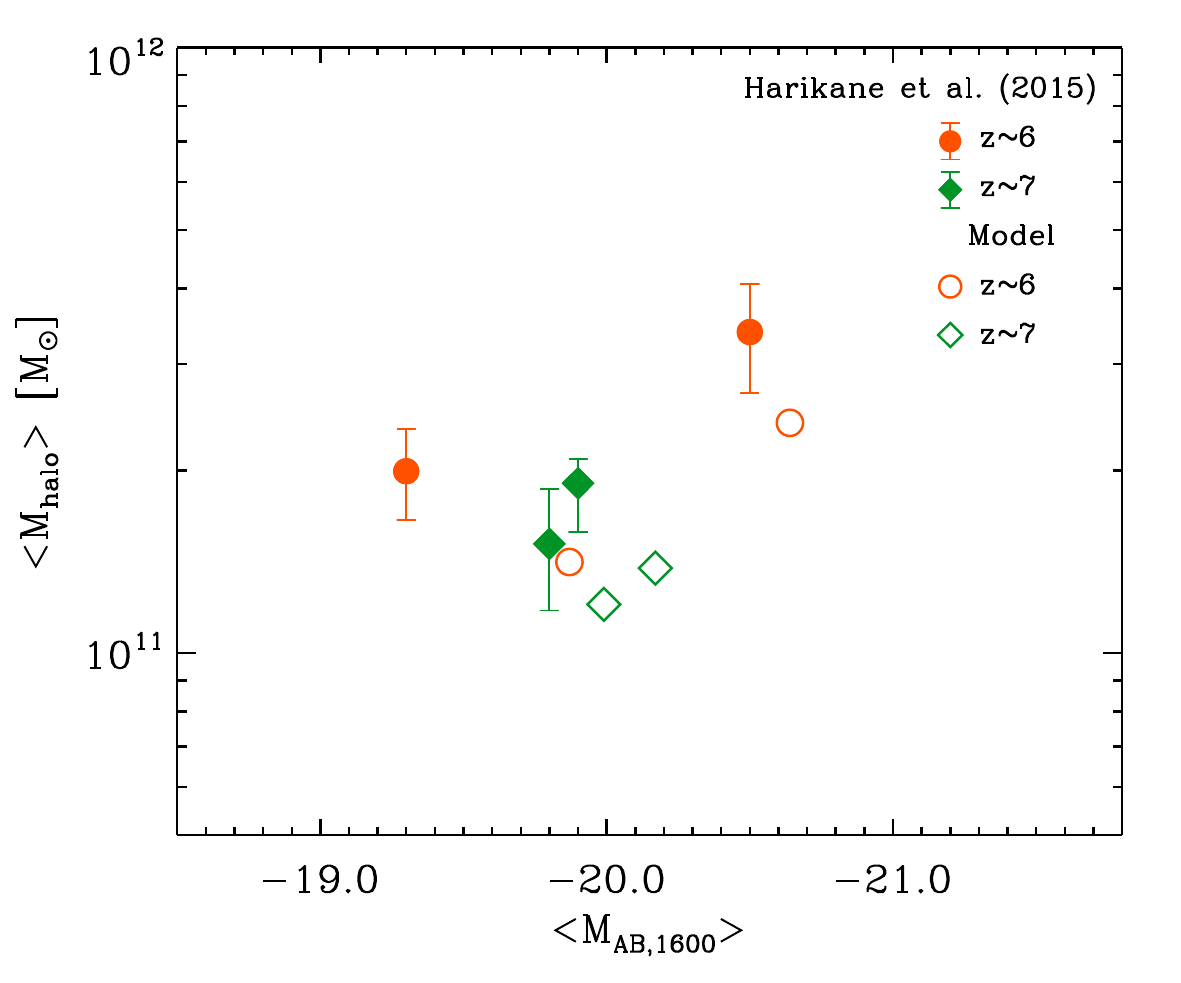}
\end{center}
\vspace{-3mm}
\caption{Mean dark matter halo mass as a function of mean UV luminosity. Filled circles and diamonds with error bars are estimated halo masses using HOD modelling from samples of \protect\cite{Harikane2015} at $z\sim6$ and $z\sim7$, respectively. Empty circles and diamonds are model predictions at $z\sim6$ and $z\sim7$, respectively.
}
\label{fig:bias3}
\end{figure}
%---------------------------------------------------------------------------------------------------------------------------------------------------%
Fig.\,\ref{fig:bias3} compares the predicted mean dark matter halo mass as a function of mean UV luminosity with the estimated halo mass from \cite{Harikane2015}. We use the same magnitude threshold for direct comparison with the observations for model LBGs (see \S.\,\ref{sec:dependence_on_luminosity}). This analysis shows that brighter LBGs reside in more massive dark matter haloes. The model predicts slightly brighter UV luminosity and less massive halo masses than the observations. However, the difference is not significant. Both observations and the model show that the host dark matter halo mass slightly decreases with increasing redshift. 

While \cite{Harikane2015} found no significant evolution of the dark matter halo mass hosting LBGs with redshift, \cite{Rob2014} reported the estimated halo mass using the combined sample from all survey fields (see Fig.\,2 in \cite{Rob2014}). They show that the dark matter halo mass hosting LBGs may be increasing from $z\sim6$ to $z\sim7$ ($\lesssim\,2\,\sigma$ confidence). On the other hand, the estimated dark matter halo masses from individual survey fields do not show this trend. \cite{Harikane2015} explained that this difference may be due to either the different sample selections or to large sample variance. Another possible difference is the method used to estimate the dark matter halo mass. \cite{Rob2014} measured the large scale bias using a variance at $8\,h^{-1}{\rm Mpc}$, and estimated the mass by calculating a mass function weighted bias. \cite{Harikane2015} estimated a mean halo mass using HOD modelling. Since the mean halo mass is affected by contribution from a range of haloes including central and satellite galaxies, the mean halo mass may be different from the halo mass estimated from large scale bias. We checked the estimated halo mass from the large scale bias of the \cite{Harikane2015} sample, and found that this shows a similar trend to that of \cite{Rob2014}, although the evolution is weaker. Confirming the evolution of dark matter halo mass hosting LBGs of fixed luminosity between $z\sim6$ and 7 will require larger surveys of LBGs.

Overall, the model prediction of galaxy bias is consistent with estimates of the observed bias, and we find that dark matter halo masses at fixed luminosity do not show significant evolution with redshift.

%%%%%%%%%%%%%%%%%%%%%%%%%%%%%%%%%%%%%%%%%%%%%%%%%%%%%%%
%                                                                                                                                                                                                  %
%      Summary and conclusions                                                                                                                                            %
%                                                                                                                                                                                                  %
%%%%%%%%%%%%%%%%%%%%%%%%%%%%%%%%%%%%%%%%%%%%%%%%%%%%%%%
\section{Conclusion}\label{sec:conclusion}
We have investigated the clustering properties of Lyman-break galaxies (LBGs) at $z\sim6$\,-\,8 using the semi-analytical model {\scshape Meraxes} developed as part of the Dark-ages Reionization And Galaxy-formation Observables from Numerical Simulation (DRAGONS) project.

This is the first study to compare model predictions using a semi-analytical model with the new clustering measurements of LBGs up to $z\sim8$. We predict the angular correlation function (ACF) of LBGs, and compare these model predictions with clustering measurements from survey fields consisting of the Hubble eXtreme Deep Field (XDF), the Hubble Ultra-Deep Field (HUDF) and Cosmic Assembly Near-infrared Deep Extragalactic Legacy Survey (CANDELS) \citep{Rob2014}, and from the HUDF and CANDELS \citep{Harikane2015}.

We find that the predicted ACFs at $z\sim 6$\,-\,8 are in good agreement with ACFs measured from observations. The model predictions show a dependence of clustering amplitude on luminosity, with brighter LBGs being more strongly clustered than fainter LBGs. The predicted dependence on luminosity is consistent with observational results. This result implies that the trend of more massive haloes hosting brighter galaxies \citep[e.g.][]{G&D2001} holds during the epoch of reionisation. Consequently, the predicted galaxy bias increases with increasing luminosity as expected in a hierarchical galaxy formation model \citep[e.g.][]{MW96}. We also find that the predicted galaxy bias at fixed apparent magnitude increases with increasing redshift, as seen in observations.

We find that the model LBGs of magnitude $M_{{\rm AB(1600)}} < -19.4$ at $6\lesssim z \lesssim 8$ reside in dark matter haloes of mean mass $\sim 10^{11.0}-10^{11.5}\,M_{\rm \odot}$, and this dark matter halo mass does not evolve significantly during reionisation. The predicted dark matter halo mass is consistent both with the estimate of the dark matter halo mass by calculating a mass function weighted bias \citep{Rob2014}, and with the estimate of the mean dark matter halo mass using HOD modelling \citep{Harikane2015}.

\vspace{5mm}

{\bf Acknowledgments} This research was supported by the Victorian Life Sciences Computation Initiative (VLSCI), grant ref. UOM0005, on its Peak Computing Facility hosted at the University of Melbourne, an initiative of the Victorian Government, Australia. Part of this work was performed on the gSTAR national facility at Swinburne University of Technology. gSTAR is funded by Swinburne and the Australian Governments Education Investment Fund. This research program is funded by the Australian Research Council through the ARC Laureate Fellowship FL110100072 awarded to JSBW. HSK is supported by a Discovery Early Career Researcher Awards (DE140100940) from the Australian Research Council. AM acknowledges support from the European Research Council (ERC) under the European Unions Horizon 2020 research and innovation program (grant agreement No 638809 AIDA). 
\newcommand{\noopsort}[1]{}

\bibliographystyle{mnras}

%\bibliography{Cross}
\bibliography{ref}

% Don't change these lines
\bsp	% typesetting comment
\label{lastpage}
\end{document}